\documentclass[%
 reprint,
 amsmath,amssymb,
aps,
]{revtex4-2}
\usepackage{graphicx}
\usepackage{dcolumn}
\usepackage{bm}
\usepackage{hyperref}
\usepackage{slashed}
\usepackage{subfloat}
\usepackage{pgfplots}
\usepackage{subfigure}
\usepackage{etoolbox}   % <-- EKLE
\usepackage{orcidlink}
\usepackage{geometry}
\definecolor{acsblue}{RGB}{17,76,139}

\begin{document}

\fontsize{7.6}{8.6}\selectfont
\preprint{APS/123-QED}

\title{Magnetic localization and spectral asymmetry of a charged particle at a screw dislocation}

\author{Claudio Furtado\orcidlink{0000-0002-3455-4285}}
\email{furtado@fisica.ufpb.br}
\affiliation{Departamento de Física, Universidade Federal da Paraíba, Caixa Postal 5008, João Pessoa, Paraíba, 58051-970, Brazil}

\author{Omar Mustafa\orcidlink{0000-0001-6664-3859}}
\email{omar.mustafa@emu.edu.tr}
\affiliation{Department of Physics, Eastern Mediterranean University, 99628, G. Magusa, north Cyprus, Mersin 10 - Türkiye}

\author{Hassan Hassanabadi\orcidlink{0000-0001-7487-6898}}
\email{hha1349@gmail.com }
\affiliation{Physics Department, California State University, Fresno, CA 93740, USA}
\affiliation{Department of Physics, Faculty of Science, University of Hradec Králové, Rokitanského 62, 500 03 Hradec Králové, Czechia}

\author{Abdullah Guvendi\orcidlink{0000-0003-0564-9899}}
\email{abdullah.guvendi@erzurum.edu.tr (corresponding author) }
\affiliation{Department of Basic Sciences, Erzurum Technical University, 25050, Erzurum, Türkiye}

\date{\today}

\begin{abstract}
{\fontsize{7.6}{8.6}\selectfont \setlength{\parindent}{0pt}
We study a charged particle moving outside a spiral (screw) dislocation in a uniform magnetic field. Although the geometry is locally Euclidean, its global screw identification couples the conserved axial and azimuthal quantities through the combination \(\mathcal{J}=\ell-\beta k\) After eliminating the cyclic coordinates, the radial motion is governed by an inverse-square term determined by $\mathcal{J}$ and a harmonic confining term generated by the magnetic field. The corresponding reduced quantum problem is exactly solvable for $q\mathcal{B}_0\neq0$, yielding an equally spaced spectrum with an orientation-dependent energy shift. The radial probability density depends on $\mathcal{J}$, and opposite values of the screw-modified quantum number have identical spatial profiles but different energies for a fixed magnetic-field orientation. The radial size scales as \(\sqrt{\langle r^2\rangle_n}\propto |q\mathcal{B}_0|^{-1/2}\), with the magnetic length setting the localization scale. At $q\mathcal{B}_0=0$, the confining term vanishes, and the problem reduces to a scale-free inverse-square system. Thus, the screw parameter enters the dynamics through the global coupling of the conserved axial and azimuthal quantities, whereas the magnetic field provides radial confinement and its orientation determines the spectral asymmetry.}
 \end{abstract}
 

\maketitle

%\tableofcontents

\section{Introduction}
\label{sec:introduction}
\setlength{\parindent}{0pt}

Geometric defects provide a natural setting for studying particle dynamics in nontrivial geometries without introducing conventional external potentials, since the effects of elastic defects can be incorporated directly into the geometric structure of the medium \cite{ref:1,ref:2}. In condensed-matter and effective-geometric descriptions, elastic distortions of a medium can be represented by an effective metric, allowing the motion of particles and quasiparticles to be formulated in terms of geometric and gauge structures \cite{ref:2,ref:3}. Dislocations are of particular interest because their geometric description need not involve local Riemann curvature: within the geometric theory of defects, dislocations are associated with torsion \cite{optic}, whereas disclinations are associated with curvature \cite{ref:1,ref:2}. A screw, or spiral, dislocation provides a simple realization of this structure; its embedding leads to a non-orthogonal metric, while the region outside the defect core is locally Euclidean and the nontrivial defect information is carried by torsion and the associated translational holonomy \cite{furtado1999landau,ref:5}. Equivalently, the Volterra construction identifies the defect through a nontrivial global matching in which an azimuthal circuit is accompanied by a translation along the defect axis \cite{ref:5,ref:6}. This distinction between local flatness and global geometric structure is directly relevant to quantum dynamics: charged particles in a screw-dislocation background exhibit modifications of magnetic spectra, scattering, and phase effects when an external magnetic field or magnetic flux is present \cite{furtado2001scattering,ref:8}.

This question has a well-established non-relativistic history. The coupling between a screw dislocation and a uniform magnetic field was first analyzed for an electron (or hole) in \cite{furtado1999landau}, where the infinite degeneracy of the Landau levels was shown to be broken by the topological boundary condition imposed by the defect; this Landau-type analysis was later extended to the simultaneous presence of a continuous distribution of disclinations, a screw dislocation, and a dispiration in \cite{marques2001landau}, and to a continuous distribution of screw dislocations, whose effective dynamics mimics that of a particle in an external magnetic field, in \cite{silvanetto2008elastic}. Complementary non-relativistic studies addressed the quantum scattering of a charged particle by an isolated screw dislocation carrying an internal magnetic flux \cite{furtado2001scattering}, the associated Aharonov--Bohm-type and Berry-phase effects \cite{furtado2000berry}, the solid-state analog of the He--McKellar--Wilkens phase for a quasiparticle moving in a charged screw-dislocation background threaded by a uniform magnetic field \cite{delimaribeiro2003hmw}, and the scattering of an induced electric dipole by a charged screw dislocation under a uniform magnetic field \cite{delimaribeiro2010scattering}. Other confining interactions combined with the screw dislocation have likewise been treated non-relativistically, including a harmonic-oscillator potential \cite{furtado2000harmonic,bueno2016harmonic}, a Tan--Inkson quantum-ring confinement in a rotating frame \cite{dantas2015ring}, an inverse-square-type interaction \cite{bakke2021inversesquare}, and an inhomogeneous magnetic field with a finite-size defect core, where a missing-flux Aharonov--Bohm effect was identified \cite{bakke2021missingflux}. The present work is set within this line of investigation, combining, in a single exactly solvable model, the screw-modified conserved momentum responsible for the geometric inverse-square barrier with the harmonic confinement generated directly by a uniform magnetic field.

The presence of a magnetic field requires that the electromagnetic structure be defined consistently on the defect geometry. In the geometric description of line defects, a screw dislocation is characterized by a nontrivial global identification associated with the Burgers vector, while its exterior geometry can remain locally flat \cite{ref:2,depaduasantos1998}. A uniform magnetic field specified in the ambient Euclidean space must therefore first be pulled back to the dislocation manifold, after which its coupling to a charged particle is determined by the intrinsic metric and minimal coupling \cite{omar-2026,guvendi-phase2026,guvendi-2-body}. The screw deformation does not modify the transverse form of the ambient gauge potential, but enters the intrinsic geometry through the screw-dependent axial one-form, thereby coupling the angular and axial momentum sectors geometrically while leaving the electromagnetic interaction inherited from the ambient field \cite{furtado2001scattering,omar-2026,guvendi-phase2026,guvendi-2-body,depaduasantos1998}. Consequently, a consistent treatment requires that the geometric embedding, global identification, gauge potential, and intrinsic field strength be specified before the particle dynamics is reduced. At the classical level, the azimuthal and axial coordinates are cyclic and therefore generate two conserved canonical momenta, but their relation is modified by the screw identification; fixing these conserved quantities reduces the dynamics to a radial degree of freedom on the exterior domain with the defect axis excluded from the configuration space. The excluded axis is physically significant because a closed circuit around the defect produces a finite axial translation determined by the Burgers vector, so global identification and associated winding sectors remain relevant even when the local Riemann curvature vanishes \cite{ref:2,depaduasantos1998}. Upon quantization, the corresponding reduced radial problem is defined on the half-line and requires a suitable boundary condition at the excluded axis together with normalizability at a large radius, making the reduced radial quantization distinct from a direct quantization of the unreduced three-dimensional system \cite{omar-2026,guvendi-phase2026,guvendi-2-body,omar-1}. In the presence of a magnetic field, the same global geometric structure is combined with the gauge coupling and produces a radial problem in which the angular and axial conserved quantities enter through a screw-dependent combination rather than as independent quantum numbers; related quantum analyses of magnetic flux in the presence of screw dislocations likewise show that the Burgers vector and electromagnetic flux can jointly affect the quantum spectrum and phase structure \cite{furtado2001scattering}. Thus, the classical Hamiltonian structure, radial motion, phase-space organization, winding sectors, and quantum dynamics can be formulated within a single geometric construction, with the effects of the dislocation arising from its intrinsic metric and global identification rather than from the introduction of an additional phenomenological potential at the defect core \cite{ref:2,depaduasantos1998}.

The remainder of this paper is organized as follows. Section~\ref{sec:spiral_geometry} introduces a Euclidean realization of the locally flat screw-dislocation geometry and establishes its intrinsic geometry, global screw identification, and Burgers vector. Section~\ref{sec:spiral_gauge} constructs the electromagnetic gauge potential by pulling back a uniform magnetic field from the ambient Euclidean space to the defect geometry and determines the corresponding intrinsic field strength. Section~\ref{sec:spiral_hamiltonian} develops the classical dynamics of the charged particle, identifies the conserved quantities, and reduces the system to the radial sector. The zero-energy condition is examined within the classical radial problem, while the phase-space structure, winding sectors, and topological properties of the spiral dislocation are analyzed in Section~\ref{sec:spiral_phase_space}. Section~\ref{sec:spiral_quantum_radial} formulates the reduced quantum problem and obtains its bound-state solutions in the presence of a nonzero magnetic field, together with a separate treatment of the zero-field sector. Finally, Section~\ref{sec:discussion} discusses the physical implications of the classical and quantum descriptions developed in the preceding sections. Throughout this work, we adopt natural units $\hbar=c=1$.

\section{Euclidean Embedding and Intrinsic Geometry of a Spiral Dislocation}
\label{sec:spiral_geometry}

\setlength{\parindent}{0pt}

The geometry of a spiral dislocation is most naturally formulated as a Euclidean embedding supplemented by a nontrivial screw identification. Let $(r,\phi,z)$ denote cylindrical coordinates on the defect exterior, with \(r>0,\, \phi\sim\phi+2\pi \), and let $\beta$ be a constant with the dimensions of length. On the universal cover of the defect exterior, the geometry admits the following Euclidean embedding
\begin{equation}
\mathbf{X}(r,\phi,z)=\left(r\cos\phi,\,r\sin\phi,\,z+\beta\phi\right).\label{eq:2.1}
\end{equation}
The parameter $\beta$ measures the axial displacement accumulated by the embedding as the angular coordinate winds around the defect axis. 
\begin{figure}[t]
\centering
\includegraphics[width=0.8\linewidth]{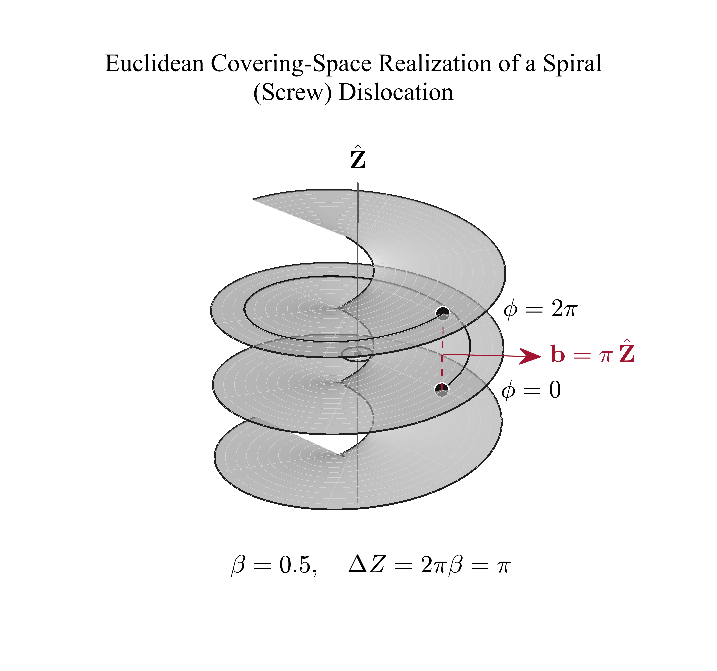}
\caption{\fontsize{7.6}{8.6}\selectfont
Three-dimensional Euclidean covering-space realization of a spiral (screw) dislocation for $\beta=0.5$, described by $\mathbf{X}(r,\phi,0)=(r\cos\phi,r\sin\phi,\beta\phi)$, with $r_{\min}=0.60$, $r_{\max}=5.50$, and $\phi\in[-3\pi,3\pi]$. The ambient Cartesian coordinates are $(X,Y,Z)$, with $X=r\cos\phi$, $Y=r\sin\phi$, and $Z=\beta\phi$. A complete $2\pi$ circuit at fixed $r$ and $z=0$ changes the ambient coordinate by $\Delta Z=2\pi\beta=\pi$, corresponding to the Burgers vector
$\mathbf{b}=2\pi\beta\,\hat{\mathbf Z}=\pi\,\hat{\mathbf Z}$. The highlighted $2\pi$ circuit has endpoints related by the screw identification and separated in the covering space by the displacement $\mathbf{b}$ along the ambient $Z$ direction. The corresponding vector marks the translational holonomy, while the vertical line denotes the ambient $Z$-axis through the excluded defect core. The figure represents the Euclidean covering-space geometry and does not introduce an additional physical spatial dimension.} \label{fig:3D}
\end{figure}
Figure~\ref{fig:3D} provides a three-dimensional Euclidean visualization of the global screw identification on the covering space, with the axial displacement $b=2\pi\beta$ representing the translational part of the defect holonomy. In the covering-space coordinates, a complete $2\pi$ circuit in the angular coordinate is accompanied by the axial shift
$\Delta Z=2\pi\beta$, so that $(r,\phi,Z)\sim(r,\phi+2\pi,Z+2\pi\beta)$. Thus, although the regular exterior is locally Euclidean, the angular identification is globally accompanied by a translation along the ambient $Z$ direction. This global screw identification provides the geometric characterization of the dislocation. The coordinate tangent vectors obtained from Eq.~\eqref{eq:2.1} are
\begin{align}
\mathbf{e}_{r}&=\partial_{r}\mathbf{X}=\left(\cos\phi,\,\sin\phi,\,0\right),\label{eq:2.2}
\\
\mathbf{e}_{\phi}&=\partial_{\phi}\mathbf{X}=\left(-r\sin\phi,\,r\cos\phi,\,\beta\right),\label{eq:2.3}
\\
\mathbf{e}_{z}&=\partial_{z}\mathbf{X}=\left(0,\,0,\,1\right).\label{eq:2.4}
\end{align}
The induced metric follows directly from the Euclidean ambient inner product \(g_{ab}=\mathbf{e}_{a}\cdot\mathbf{e}_{b}\), and is therefore
\begin{equation}
\begin{split}
ds^{2}&=dr^{2}+\left(r^{2}+\beta^{2}\right)d\phi^{2}+2\beta\,d\phi\,dz+dz^{2}\Rightarrow\\
ds^{2}&=dr^{2}+r^{2}d\phi^{2}+\left(dz+\beta\,d\phi\right)^{2}.\label{eq:2.6}
\end{split}
\end{equation}
This form makes the geometric content particularly transparent. The screw parameter does not modify the radial line element or the coefficient of the ordinary azimuthal term; instead, it mixes the angular and axial directions through the globally nontrivial one-form $dz+\beta\,d\phi$. The local geometry is consequently Euclidean in a suitably adapted coordinate system, whereas the global identification
retains the dislocation. In the coordinate basis $(r,\phi,z)$, the metric and its inverse are \cite{omar-1}
\begin{equation}
g_{ab}
=
\begin{pmatrix}
1 & 0 & 0\\
0 & r^{2}+\beta^{2} & \beta\\
0 & \beta & 1
\end{pmatrix},
\qquad
g^{ab}
=
\begin{pmatrix}
1 & 0 & 0\\
0 & \displaystyle\frac{1}{r^{2}}
&
\displaystyle-\frac{\beta}{r^{2}}
\\[6pt]
0
&
\displaystyle-\frac{\beta}{r^{2}}
&
\displaystyle1+\frac{\beta^{2}}{r^{2}}
\end{pmatrix}.
\label{eq:2.7}
\end{equation}
The volume density $g$ is \cite{omar-1}
\begin{equation}
g=\det(g_{ab})=r^{2}\quad\Rightarrow\quad \sqrt{g}=r.\label{eq:2.8}
\end{equation}
The inverse metric in Eq.~\eqref{eq:2.7} exhibits explicitly the angular-axial mixing generated by the screw parameter. At the same time, the determinant remains exactly $r^{2}$, independent of $\beta$. Hence the screw deformation changes the inverse metric without modifying the coordinate volume density. A useful way of exposing the local character of this geometry is to introduce, on any simply connected patch that does not intersect the defect axis, the adapted coordinate \(Z=z+\beta\phi\). Since \(dZ=dz+\beta\,d\phi\), Eq.~\eqref{eq:2.6} becomes locally
\begin{equation}
ds^{2}=dr^{2}+r^{2}d\phi^{2}+dZ^{2}.\label{eq:2.12}
\end{equation}
Thus, on every regular simply connected patch with $r>0$, the metric is locally identical to the Euclidean metric written in cylindrical coordinates. It does not, however, trivialize the global defect, because the coordinate transformation \(Z=z+\beta\phi\) is not globally single-valued under a complete azimuthal circuit. Indeed, under \(\phi\to\phi+2\pi\), the adapted coordinate \(Z=z+\beta\phi\) transforms as \(Z\to Z+2\pi\beta\). Thus the standard cylindrical identification \((r,\phi,z)\sim(r,\phi+2\pi,z)\) is equivalently expressed in the locally Euclidean coordinates \((r,\phi,Z)\) as the screw identification
\begin{equation}
(r,\phi,Z)\sim(r,\phi+2\pi,Z+2\pi\beta).
\label{eq:2.14}
\end{equation}
For a particle of charge \(q\), the electromagnetic gauge potential is the one-form
\begin{equation}
A=A_\mu\,dx^\mu=A_r\,dr+A_\phi\,d\phi+A_Z\,dZ
\end{equation}
and the corresponding gauge holonomy around a closed azimuthal loop \(C\) with \(\phi:0\to2\pi\) is
\begin{equation}
\mathcal{U}(C)=\exp\!\left(iq\oint_C A\right).
\end{equation}
The local flattening transformation and the global quotient must therefore be distinguished sharply. Local curvature detects only the differential geometry of the regular exterior, whereas the screw identification retains information that cannot be removed by a single globally defined smooth coordinate transformation \cite{ref:1}. The Levi-Civita connection associated with the induced metric is nontrivial. The Christoffel symbols are obtained from \cite{ref:1}
\begin{equation}
\Gamma^{a}{}_{bc}
=
\frac{1}{2}g^{ad}
\left(
\partial_{b}g_{dc}
+
\partial_{c}g_{db}
-
\partial_{d}g_{bc}
\right).
\label{eq:2.15}
\end{equation}
Since the metric coefficients depend only on $r$, the only nonvanishing connection coefficients for $r>0$ are
\begin{align}
\Gamma^{r}{}_{\phi\phi}
&=
-r,
&
\Gamma^{\phi}{}_{r\phi}
&=
\Gamma^{\phi}{}_{\phi r}
=
\frac{1}{r},
&
\Gamma^{z}{}_{r\phi}
&=
\Gamma^{z}{}_{\phi r}
=
-\frac{\beta}{r}.
\label{eq:2.19}
\end{align}
The first two coefficients are the familiar connection terms associated with cylindrical coordinates. The last coefficient is generated directly by the screw-induced off-diagonal component $g_{\phi z}=\beta$. The nonzero Christoffel symbols do not imply a curved regular manifold. They characterize parallel transport in the chosen coordinate basis. The invariant local curvature is instead determined by the Riemann tensor \cite{guvendi2026charged},
\begin{equation}
R^{a}{}_{bcd}=\partial_{c}\Gamma^{a}{}_{db}-\partial_{d}\Gamma^{a}{}_{cb}+\Gamma^{a}{}_{ce}\Gamma^{e}{}_{db}-\Gamma^{a}{}_{de}\Gamma^{e}{}_{cb}.\label{eq:2.18}
\end{equation}
Accordingly, one obtains For the radial--angular sector one obtains
\begin{align}
R^{\phi}{}_{r\phi r}=0,\qquad R^{r}{}_{\phi r\phi}=0,\qquad R^{z}{}_{r\phi r}=0, \qquad
r>0\label{eq:2.20}
\end{align}
The regular spiral-dislocation exterior is therefore locally Euclidean. In particular, there is no nonzero smooth curvature invariant proportional to $\beta$ in the regular region $r>0$. The vanishing of the regular Riemann tensor must not, however, be interpreted as the absence of the dislocation itself. The defect is encoded in the global screw identification \eqref{eq:2.14}. The geometry is therefore locally flat but globally nontrivial. This distinction may also be expressed in terms of parallel transport. For any contractible loop contained entirely in the regular region, the vanishing of the Riemann tensor implies trivial local curvature holonomy. A loop that winds nontrivially around the excluded axis, however, probes the global quotient structure and retains the screw translation $2\pi\beta$. Thus the defect information is topological or global rather than encoded in a smooth local curvature invariant. The global content can be characterized by the corresponding Burgers vector \cite{ref:1,ref:2}. For one complete winding around the defect axis, the axial displacement is
\begin{equation}
b_{z}=\oint \beta\,d\phi=2\pi\beta \Rightarrow \mathbf{b}=2\pi\beta\,\hat{\mathbf z}.\label{eq:2.35}
\end{equation}
Hence the Burgers vector $\mathbf{b}$ is directed along the screw axis. This quantity is invariantly associated with the global translational part of the screw holonomy. It survives in the locally flat description and therefore provides a direct measure of the defect that
is not reducible to the regular scalar curvature. 

\section{Gauge Pullback and Electromagnetic Structure of the Spiral Dislocation}
\label{sec:spiral_gauge}

\setlength{\parindent}{0pt}

\begin{figure}[t]
\centering
\includegraphics[width=0.8\linewidth]{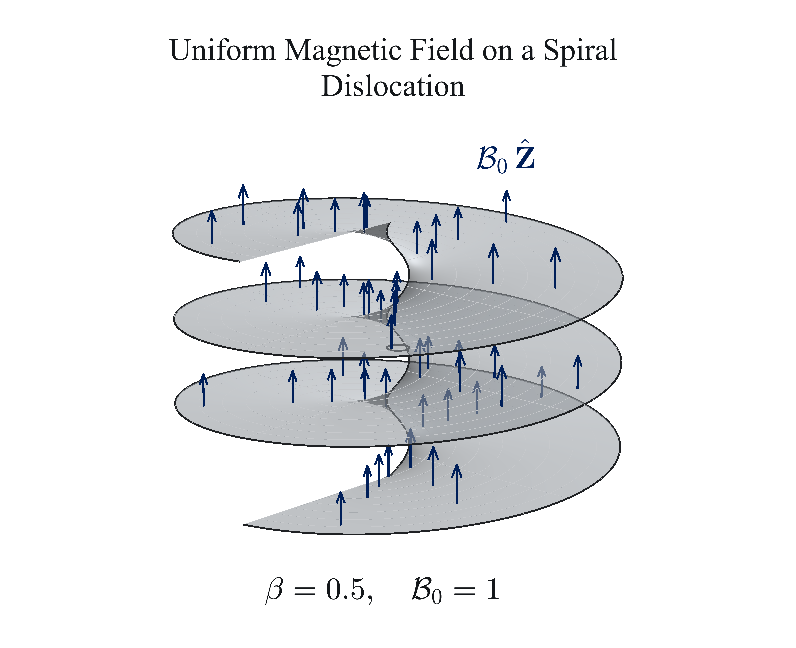}
\caption{\fontsize{7.6}{8.6}\selectfont
Euclidean covering-space realization of the spiral (screw) dislocation, $\mathbf{X}(r,\phi,0)=(r\cos\phi,r\sin\phi,\beta\phi)$, with $\beta=0.5$, $r_{\min}=0.60$, and $r_{\max}=8.0$, in a uniform ambient magnetic field $\boldsymbol{\mathcal B}_{0}=\mathcal{B}_{0}\hat{\mathbf Z}$ with $\mathcal{B}_{0}=1$. The navy-blue arrows indicate the ambient magnetic-field direction, everywhere parallel to the ambient $Z$ axis. The pulled-back field has coordinate component $\mathcal{F}_{r\phi}=\mathcal{B}_{0}r$ and locally measured magnitude $|\boldsymbol{\mathcal B}|=|\mathcal{B}_{0}|$, while the screw parameter $\beta$ enters the intrinsic geometry through the metric and the global screw identification.}
\label{fig:MF}
\end{figure}

Having established the Euclidean embedding and intrinsic geometry of the spiral dislocation in Sec.~\ref{sec:spiral_geometry}, we now determine the electromagnetic field induced on the defect by a uniform magnetic field in the embedding space (see also \cite{omar-2026,guvendi-phase2026,guvendi-2-body}). The construction is performed by pulling back the ambient electromagnetic gauge potential to the spiral-dislocation geometry. Throughout this section, $(X,Y,Z)$ denote Cartesian coordinates of the three-dimensional Euclidean embedding space, while $(r,\phi,z)$ denote the intrinsic cylindrical coordinates, related by Eq. \ref{eq:2.1}. Thus $z$ is an independent intrinsic coordinate, while the screw deformation is encoded in the shifted ambient coordinate $Z$. Locally, and on the Euclidean covering space, this relation provides the corresponding embedding coordinates. Globally, the azimuthal identification $\phi\sim\phi+2\pi$ is accompanied by the screw translation $Z\sim Z+2\pi\beta$. For a uniform magnetic field $\mathcal{B}_{0}$ directed along the positive $Z$ axis (see Figure \ref{fig:MF}), we choose the symmetric gauge
\begin{equation}
\boldsymbol{\mathcal A}_{\mathrm{amb}}=\frac{\mathcal B_{0}}{2}\left(X\,dY-Y\,dX\right),\label{eq:3.2}
\end{equation}
for which \(d\boldsymbol{\mathcal A}_{\mathrm{amb}}=\mathcal B_{0}\,dX\wedge dY\).
Let
\begin{equation}
\iota:(r,\phi,z)\longmapsto\left(r\cos\phi,\,r\sin\phi,\,z+\beta\phi\right)\label{eq:3.5}
\end{equation}
Locally, $\iota$ provides the Euclidean realization of the intrinsic geometry. Globally, the map is understood on the covering space, with the screw identification $(r,\phi,Z)\sim(r,\phi+2\pi,Z+2\pi\beta)$. Since
\begin{equation}
X\,dY-Y\,dX=r^{2}\,d\phi ,
\end{equation}
the pullback of the ambient gauge potential is
\begin{equation}
\mathcal A=\iota^{*}\boldsymbol{\mathcal A}_{\mathrm{amb}}=\frac{\mathcal B_{0}}{2}r^{2}\,d\phi .
\end{equation}
Hence its intrinsic components are
\begin{equation}
\mathcal A_{r}=0,\qquad\mathcal A_{\phi}=\frac{\mathcal B_{0}}{2}r^{2},\qquad\mathcal A_{z}=0 .
\label{eq:3.6}
\end{equation}
The absence of an explicit $\beta$ dependence in $\mathcal A_{\phi}$ is a direct consequence of the fact that the chosen ambient gauge depends only on the transverse Cartesian coordinates $X$ and $Y$. The screw parameter instead enters the intrinsic geometry through
\begin{equation}
dZ=dz+\beta\,d\phi .
\end{equation}
Consequently, although the pulled-back gauge potential has no explicit $\beta$ dependence in this gauge, the charged-particle dynamics acquires $\beta$ dependence through the intrinsic metric and its inverse. The intrinsic field strength follows from the pullback of the ambient curvature two-form,
\begin{equation}
\mathcal F=d\mathcal A=\iota^{*}\left(d\boldsymbol{\mathcal A}_{\mathrm{amb}}\right).
\end{equation}
Using
\begin{equation}
d\mathcal A=\frac{\mathcal B_{0}}{2}d(r^{2})\wedge d\phi ,
\end{equation}
we obtain
\begin{equation}
\mathcal F=\mathcal B_{0}r\,dr\wedge d\phi .
\label{eq:3.7}
\end{equation}
Therefore, the nonvanishing coordinate components are
\begin{equation}
\mathcal F_{r\phi}=\mathcal B_{0}r,\qquad\mathcal F_{\phi r}=-\mathcal B_{0}r.
\label{eq:3.8}
\end{equation}
The intrinsic metric obtained from the Euclidean embedding is given in Eq. \eqref{eq:2.6}. It is therefore natural to introduce the orthonormal coframe
\begin{equation}
\vartheta^{1}=dr,\qquad\vartheta^{2}=r\,d\phi,\qquad\vartheta^{3}=dz+\beta\,d\phi ,\label{eq:3.10a}
\end{equation}
for which \(d\ell^{2}=(\vartheta^{1})^{2}+(\vartheta^{2})^{2}+(\vartheta^{3})^{2}\). Accordingly, the corresponding dual orthonormal frame is
\begin{equation}
\mathbf e_{1}=\partial_{r},\qquad
\mathbf e_{2}
=\frac{1}{r}\left(\partial_{\phi}-\beta\,\partial_{z}\right),
\qquad
\mathbf e_{3}=\partial_{z}.\label{eq:3.10b}
\end{equation}
Since
\begin{equation}
dr\wedge d\phi=\frac{1}{r}\vartheta^{1}\wedge\vartheta^{2},
\end{equation}
the electromagnetic two-form takes the particularly simple form
\begin{equation}
\mathcal F=\mathcal B_{0}\,\vartheta^{1}\wedge\vartheta^{2}.\label{eq:3.11}
\end{equation}
Thus the magnetic field measured by a local orthonormal observer is
\begin{equation}
\boldsymbol{\mathcal B}=\mathcal B_{0}\,\mathbf e_{3},\qquad
\left|\boldsymbol{\mathcal B}\right|
=\left|\mathcal B_{0}\right|,\label{eq:3.12}
\end{equation}
where $\mathbf e_{3}$ is the vector dual to $\vartheta^{3}$. The factor of $r$ in the coordinate component $\mathcal F_{r\phi}=\mathcal B_{0}r$ arises from the use of the coordinate
one-form $d\phi$, whose proper orthonormal normalization is $r\,d\phi$. In the orthonormal frame, the field strength is simply $\mathcal F=\mathcal B_{0}\vartheta^{1}\wedge\vartheta^{2}$, so that the
locally measured magnetic-field magnitude is $|\mathcal B_{0}|$. Importantly, the screw parameter $\beta$ does not alter the local magnitude of the imposed magnetic field. Instead, it modifies the relation between the coordinate and orthonormal angular directions, as seen explicitly from \(\mathbf e_{2}=\frac{1}{r}\left(\partial_{\phi}-\beta\,\partial_{z}\right)\), while \(\mathbf e_{3}=\partial_{z}\) remains the unit longitudinal vector. The $\beta$ dependence consequently enters the charged-particle dynamics through the intrinsic metric and its inverse, rather than through a renormalization of $\mathcal B_{0}$. The gauge character of the construction follows directly from the pullback. Under an ambient Abelian gauge transformation,
\begin{equation}
\boldsymbol{\mathcal A}_{\mathrm{amb}}\longrightarrow\boldsymbol{\mathcal A}_{\mathrm{amb}}+d\Lambda ,
\end{equation}
the intrinsic gauge potential transforms as
\begin{equation}
\mathcal A\longrightarrow\mathcal A+d\lambda,\qquad\lambda=\iota^{*}\Lambda .\label{eq:3.13}
\end{equation}
The corresponding field strength is invariant:
\begin{align}
\mathcal F\longrightarrow d\left(\mathcal A+d\lambda\right)=d\mathcal A+d^{2}\lambda=\mathcal F,\qquad d^{2}=0 .\label{eq:3.14}
\end{align}
Hence $\mathcal A_{\phi}$ in Eq.~\eqref{eq:3.6} is gauge dependent and enters the charged-particle Hamiltonian through minimal coupling, whereas $\mathcal F$ in Eq.~\eqref{eq:3.11} is gauge invariant and represents the physical electromagnetic field. Finally, the global screw character of the geometry is encoded in the translational holonomy associated with an azimuthal circuit. For a loop at
fixed $r$ and $z$, the curve is closed in the physical quotient geometry, whereas its lift to the Euclidean covering space acquires an axial displacement. Indeed,
\begin{equation}
\oint\left(dZ-dz\right)
=\oint\beta\,d\phi
=2\pi\beta,
\end{equation}
so that the corresponding Burgers vector is
\begin{equation}
\mathbf b
=2\pi\beta\,\hat{\mathbf Z}.
\end{equation}
Thus the translational part of the screw holonomy survives despite the vanishing of the regular Riemann curvature on the defect exterior. It provides a global characterization of the dislocation that is not captured by local scalar curvature invariants. This quantity is invariantly associated with the global translational part of the screw holonomy. It survives in the locally flat description and therefore provides a direct measure of the defect that is not reducible to the regular scalar curvature. Together, the gauge potential in Eq.~\eqref{eq:3.6}, its gauge-invariant field strength in Eq.~\eqref{eq:3.11}, and the intrinsic metric in Eq.~\eqref{eq:2.6} provide the complete electromagnetic and geometric input
for the charged-particle dynamics developed below.

\section{Charged Particle Dynamics on the Spiral-dislocation Geometry}
\label{sec:spiral_hamiltonian}

\setlength{\parindent}{0pt}

The intrinsic metric and the pulled-back electromagnetic one-form obtained in Secs.~\ref{sec:spiral_geometry} and \ref{sec:spiral_gauge} completely determine the classical motion of a charged particle on the regular spiral-dislocation exterior. Since the Levi--Civita Riemann tensor vanishes identically for $r>0$, the regular geometry does not generate an additional local
curvature-induced force or an independent curvature potential. The mechanical effect of the dislocation instead originates from the non-orthogonal $(\phi,z)$ sector of the metric, whose off-diagonal component
$g_{\phi z}=\beta$ couples the azimuthal and axial motions. In the presence of the ambient uniform magnetic field, the electromagnetic interaction is fixed by the pullback derived in Sec.~\ref{sec:spiral_gauge}; consequently, no additional $\beta$-dependent contribution is to be introduced into the electromagnetic potential. The screw parameter therefore enters the dynamics
geometrically, through the momentum coupling generated by the intrinsic metric, whereas the magnetic interaction is determined by the ambient field strength and its pullback.

For a particle of mass $\mu$ and charge $q$, the minimally coupled Lagrangian \(\mathcal{L}
=\frac{\mu}{2}\,g_{ij}\dot{x}^{i}\dot{x}^{j}+q\,\mathcal{A}_{i}\dot{x}^{i}\) on the regular region $r>0$ is \cite{omar-2026,guvendi-phase2026,guvendi-2-body}
\begin{equation}
\mathcal{L}
=
\frac{\mu}{2}
\left[
\dot r^{\,2}
+
r^{2}\dot\phi^{\,2}
+
\left(\dot z+\beta\dot\phi\right)^{2}
\right]
+
q\mathcal{A}_{i}\dot x^{i},
\label{eq:4.3}
\end{equation}
where, from Eq.~\eqref{eq:3.6},
\begin{equation}
q\mathcal{A}_{i}\dot x^{i}
=
\frac{q\mathcal{B}_{0}}{2}r^{2}\dot\phi.
\label{eq:4.4}
\end{equation}
Hence, the explicit Lagrangian reads
\begin{equation}
\mathcal{L}
=
\frac{\mu}{2}
\left[
\dot r^{\,2}
+
r^{2}\dot\phi^{\,2}
+
\left(\dot z+\beta\dot\phi\right)^{2}
\right]
+
\frac{q\mathcal{B}_{0}}{2}r^{2}\dot\phi.
\label{eq:4.5}
\end{equation}
The magnetic term contains no explicit $\beta$ dependence because the ambient symmetric gauge depends only on the transverse coordinates $X$ and $Y$, as established in Sec.~\ref{sec:spiral_gauge}. The screw parameter enters the particle dynamics through the intrinsic kinetic term $(\dot z+\beta\dot\phi)^2$. The canonical momenta are obtained directly from Eq.~\eqref{eq:4.5} \cite{omar-2026,guvendi-phase2026,guvendi-2-body}:
\begin{align}
p_{r}
&=
\frac{\partial\mathcal{L}}{\partial\dot r}
=
\mu\dot r,
\label{eq:4.6a}
\\
p_{\phi}
&=
\frac{\partial\mathcal{L}}{\partial\dot\phi}
=
\mu r^{2}\dot\phi
+
\mu\beta\left(\dot z+\beta\dot\phi\right)
+
\frac{q\mathcal{B}_{0}}{2}r^{2},
\label{eq:4.6b}
\\
p_{z}
&=
\frac{\partial\mathcal{L}}{\partial\dot z}
=
\mu\left(\dot z+\beta\dot\phi\right).
\label{eq:4.6c}
\end{align}
Because neither $\phi$ nor $z$ occurs explicitly in the Lagrangian, both are cyclic coordinates. Their conjugate canonical momenta are therefore conserved:
\begin{equation}
p_{\phi}=\ell,
\qquad
p_{z}=k,
\label{eq:4.7}
\end{equation}
where $\ell$ denotes the conserved canonical azimuthal momentum and $k$ the conserved momentum conjugate to the intrinsic axial coordinate $z$. Solving Eq.~\eqref{eq:4.6c} for the axial velocity gives
\begin{equation}
\dot z+\beta\dot\phi
=
\frac{k}{\mu},
\label{eq:4.8}
\end{equation}
or, equivalently,
\begin{equation}
\dot z
=
\frac{k}{\mu}
-
\beta\dot\phi.
\label{eq:4.9}
\end{equation}
The latter relation makes explicit that the coordinate velocity $\dot z$ is not, in general, equal to the axial mechanical velocity measured along the orthonormal direction dual to $\vartheta^{3}=dz+\beta\,d\phi$. The combination $\dot z+\beta\dot\phi$, rather than $\dot z$ alone, is the geometrically natural longitudinal velocity.

It is useful to introduce the gauge-shifted azimuthal momentum
\begin{equation}
\Pi_{\phi}
\equiv
p_{\phi}-q\mathcal{A}_{\phi}
=
p_{\phi}
-
\frac{q\mathcal{B}_{0}}{2}r^{2}.
\label{eq:4.10}
\end{equation}
Using Eq.~\eqref{eq:4.6b} together with $p_z=k$, one obtains
\begin{equation}
\Pi_{\phi}
=
\mu r^{2}\dot\phi
+
\beta p_{z},
\label{eq:4.11}
\end{equation}
and therefore
\begin{equation}
\dot\phi
=
\frac{\Pi_{\phi}-\beta p_{z}}
{\mu r^{2}}.
\label{eq:4.12}
\end{equation}
Thus, the quantity that governs the azimuthal mechanical motion is not the canonical momentum $p_{\phi}$ by itself, but the combination $\Pi_{\phi}-\beta p_{z}$. This is the direct dynamical manifestation of the off-diagonal metric component $g_{\phi z}=\beta$: the screw geometry converts a portion of the conserved axial momentum into an effective shift of the azimuthal mechanical momentum.

The Hamiltonian is obtained by the Legendre transformation \cite{omar-2026,guvendi-phase2026,guvendi-2-body}
\begin{equation}
\mathcal{H}
=
p_{r}\dot r
+
p_{\phi}\dot\phi
+
p_{z}\dot z
-
\mathcal{L}.
\label{eq:4.13}
\end{equation}
Using the inverse metric in \eqref{eq:2.7}, and the minimally coupled momentum
$\Pi_{\phi}=p_{\phi}-q\mathcal{A}_{\phi}$, this Hamiltonian can be written as
\begin{align}
\mathcal{H}=\frac{1}{2\mu}\left[p_{r}^{\,2}+p_{z}^{\,2}+\frac{\left(\Pi_{\phi}-\beta p_{z}\right)^{2}}{r^{2}}\right].\label{eq:4.14}
\end{align}
In particular, the screw deformation does not generate an additional electromagnetic interaction:
its contribution arises entirely from the inverse intrinsic metric. The magnetic field enters through the gauge-covariant momentum $\Pi_{\phi}=p_{\phi}-q\mathcal{A}_{\phi}$, while the $\beta$ dependence arises from the geometric coupling between the azimuthal and axial momentum sectors. Upon using the conserved quantities $p_{\phi}=\ell$ and $p_{z}=k$, Eq.~\eqref{eq:4.10} gives
\begin{equation}
\Pi_{\phi}
=
\ell
-
\frac{q\mathcal{B}_{0}}{2}r^{2},
\label{eq:4.15}
\end{equation}
and the Hamiltonian becomes
\begin{equation}
\mathcal{H}
=
\frac{k^{2}}{2\mu}
+
\frac{p_{r}^{\,2}}{2\mu}
+
\frac{
\left[
\ell-\beta k
-
\dfrac{q\mathcal{B}_{0}}{2}r^{2}
\right]^{2}
}
{2\mu r^{2}}.
\label{eq:4.16}
\end{equation}
The first term is the conserved axial kinetic contribution, whereas the
remaining terms describe the reduced radial dynamics. It is therefore
natural to define
\begin{equation}
\mathcal{J}\equiv\ell-\beta k,
\qquad
a\equiv\frac{q\mathcal{B}_{0}}{2},
\label{eq:4.17}
\end{equation}
so that
\begin{equation}
\mathcal{H}
=
\frac{k^{2}}{2\mu}
+
\frac{p_{r}^{\,2}}{2\mu}
+
\mathcal{V}_{\mathrm{eff}}(r),
\label{eq:4.18}
\end{equation}
with
\begin{equation}
\mathcal{V}_{\mathrm{eff}}(r)
=
\frac{(\mathcal{J}-ar^{2})^{2}}{2\mu r^{2}}
=
\frac{\mathcal{J}^{2}}{2\mu r^{2}}
-
\frac{a\mathcal{J}}{\mu}
+
\frac{a^{2}r^{2}}{2\mu}.
\label{eq:4.19}
\end{equation}
This decomposition separates the three distinct contributions to the radial potential. The first is the inverse-square term generated by the screw-shifted angular momentum $\mathcal{J}$, the second is a constant magnetic cross-term, and the third is the quadratic term responsible for magnetic confinement at large radius. The screw parameter $\beta$ therefore enters the radial problem only through the shifted conserved momentum $\mathcal{J}$, while the magnetic scale is determined by the ambient field through $a=q\mathcal{B}_{0}/2$. Writing the conserved total energy as $\mathcal{E}$ and separating the
constant axial contribution gives
\begin{equation}
\mathcal{E}_{\perp}
\equiv
\mathcal{E}
-
\frac{k^{2}}{2\mu},
\label{eq:4.20}
\end{equation}
so that the radial Hamiltonian takes the one-dimensional form
\begin{equation}
\mathcal{E}_{\perp}
=
\frac{p_{r}^{\,2}}{2\mu}
+
\mathcal{V}_{\mathrm{eff}}(r).
\label{eq:4.21}
\end{equation}
The radial turning points are consequently determined by $p_r=0$, or
equivalently,
\begin{equation}
\mathcal{V}_{\mathrm{eff}}(r_t)
=
\mathcal{E}_{\perp}.
\label{eq:4.22}
\end{equation}
Introducing $y=r_t^{2}$, the turning-point equation becomes
\begin{equation}
a^{2}y^{2}
-
2\left(
\mu\mathcal{E}_{\perp}+a\mathcal{J}
\right)y
+
\mathcal{J}^{2}
=
0.
\label{eq:4.23}
\end{equation}
For $a\neq0$, its two algebraic roots are
\begin{equation}
r_{\pm}^{\,2}
=
\frac{
\mu\mathcal{E}_{\perp}
+
a\mathcal{J}
\pm
\sqrt{
\mu\mathcal{E}_{\perp}
\left(
\mu\mathcal{E}_{\perp}+2a\mathcal{J}
\right)
}
}
{a^{2}}.
\label{eq:4.24}
\end{equation}
The product of the two turning-point radii is independent of the energy:
\begin{equation}
r_{+}r_{-}=\frac{|\mathcal{J}|}{|a|}.\label{eq:4.29}
\end{equation}
This energy-independent relation follows directly from the quadratic structure in $r^{2}$ and shows that the product of the two turning radii is fixed entirely by the competition between the magnetic scale $|a|$ and the screw-shifted angular momentum $|\mathcal{J}|$. Equivalently, their geometric mean is
\begin{equation}
\sqrt{r_{+}r_{-}}=\sqrt{\frac{|\mathcal{J}|}{|a|}}.
\label{eq:4.29-a}
\end{equation}
The curvature of the reduced radial potential at the stationary point is
\begin{equation}
\left.
\frac{d^{2}\mathcal{V}_{\mathrm{eff}}}{dr^{2}}
\right|_{r=r_{\mathrm{c}}}
=
\frac{1}{\mu}
\left(
\frac{3\mathcal{J}^{2}}{r_{\mathrm{c}}^{4}}
+
a^{2}
\right)
>0,
\label{eq:4.32}
\end{equation}
for $\mu>0$ and $a,\mathcal{J}\neq0$. Hence, the stationary point is not merely a local extremum but the unique strict global minimum of the effective radial potential on the regular domain $r>0$. Evaluating
Eq.~\eqref{eq:4.19} at $r=r_{\mathrm{c}}$ gives
\begin{equation}
\mathcal{V}_{\min}
=
\frac{|a\mathcal{J}|-a\mathcal{J}}{\mu}.
\label{eq:4.33}
\end{equation}
The minimum therefore depends qualitatively on the relative sign of the
magnetic scale $a$ and the screw-shifted angular momentum $\mathcal{J}$. More
explicitly,
\begin{equation}
\begin{cases}
a\mathcal{J}>0:
&
r_{\mathrm{c}}^{2}
=
\dfrac{\mathcal{J}}{a},
\qquad
\mathcal{V}_{\min}=0,
\\[10pt]
a\mathcal{J}<0:
&
r_{\mathrm{c}}^{2}
=
-\dfrac{\mathcal{J}}{a},
\qquad
\mathcal{V}_{\min}
=
-\dfrac{2a\mathcal{J}}{\mu}>0.
\end{cases}
\label{eq:4.34}
\end{equation}
Equivalently, the equilibrium radius can be written in the sign-independent form
\begin{equation}
r_{\mathrm{c}}^{2}
=
\frac{|\mathcal{J}|}{|a|},
\qquad
r_{\mathrm{c}}
=
\sqrt{\frac{|\mathcal{J}|}{|a|}}.
\end{equation}
For $a\mathcal{J}>0$, the two terms entering the squared numerator of Eq.~\eqref{eq:4.19} can cancel exactly at the equilibrium radius,
\begin{equation}
\mathcal{J}-ar_{\mathrm{c}}^{2}=0,
\end{equation}
so that the reduced radial potential attains its absolute minimum,
$\mathcal{V}_{\min}=0$. In contrast, when $a\mathcal{J}<0$, the quantities $\mathcal{J}$ and
$ar^{2}$ necessarily have opposite signs for every $r>0$; consequently,
the condition $\mathcal{J}-ar^{2}=0$ has no physical solution. Exact cancellation
is then impossible, and the radial potential remains strictly positive,
with the nonzero minimum $\mathcal{V}_{\min}=-2a\mathcal{J}/\mu>0$. The sign of $a\mathcal{J}$ thus determines whether the magnetic coupling can completely compensate the screw-shifted
canonical angular momentum or instead leaves an irreducible positive
radial-energy offset. The global character of this minimum is further clarified by the two
asymptotic sectors of the effective potential. Directly from
Eq.~\eqref{eq:4.19}, the behavior near the excluded defect axis is
\begin{equation}
\mathcal{V}_{\mathrm{eff}}(r)
\sim
\frac{\mathcal{J}^{2}}{2\mu r^{2}},
\qquad
r\rightarrow0^{+},
\qquad
\mathcal{J}\neq0,
\label{eq:4.35}
\end{equation}
whereas at large radius one finds
\begin{equation}
\mathcal{V}_{\mathrm{eff}}(r)
\sim
\frac{a^{2}r^{2}}{2\mu},
\qquad
r\rightarrow\infty,
\qquad
a\neq0.
\label{eq:4.36}
\end{equation}
Thus, for $\mathcal{J}\neq0$, the effective potential diverges as
$r^{-2}$ on approaching the excluded defect axis, while for
$a\neq0$ it grows quadratically at large radius. These two asymptotic
limits independently establish the radial confinement: the inverse-square
sector prevents the trajectory from reaching the singular axis, whereas
the magnetic quadratic sector prevents unbounded radial escape. Together
with the strict positivity of Eq.~\eqref{eq:4.32}, they demonstrate that
the stationary point identified above is the unique global minimum of the
radial dynamics on $r>0$. The two asymptotic scales have distinct physical origins. The
inverse-square barrier is controlled entirely by the conserved
screw-shifted momentum \(\mathcal{J}\) so that the screw parameter $\beta$ modifies the strength of the
short-distance barrier through the conserved combination of azimuthal and
axial momenta, while leaving its universal $r^{-2}$ scaling unchanged.
By contrast, the large-$r$ confinement is governed by $a=\frac{q\mathcal{B}_{0}}{2}$, and therefore originates directly from the ambient uniform magnetic field.
The magnetic field fixes the coefficient of the asymptotic harmonic term
without changing the local $r\rightarrow0^{+}$ scaling of the geometric
centrifugal sector. The radial problem consequently exhibits a clean
separation between the defect-induced momentum shift, encoded in $\mathcal{J}$, and
the electromagnetic confinement scale, encoded in $a$. This structure also makes transparent why the energy-independent turning point relation obtained in Eq.~\eqref{eq:4.29} is possible. The inverse-
square divergence at small $r$ and the quadratic growth at large $r$ combine to produce a radial potential with a single finite minimum. Consequently, for energies above the minimum, the radial motion is bounded between two positive turning radii whenever the corresponding roots of the turning-point equation are real and positive. Their product remains fixed by the coefficients of the quadratic polynomial in $r^{2}$ and is therefore independent of the particle energy. The energy changes the
location of the two turning points individually, but not their product, which is controlled solely by the competition between the magnetic scale $|a|$ and the screw-shifted angular-momentum scale $|\mathcal{J}|$.

Finally, the axial coordinate remains cyclic. The corresponding conserved
canonical momentum is
\begin{equation}
p_{z}
=
\frac{\partial\mathcal{L}}{\partial\dot z}
=
\mu\left(\dot z+\beta\dot\phi\right)
+
q\mathcal{A}_{z}.
\label{eq:4.38}
\end{equation}
For the gauge choice used here, $\mathcal{A}_{z}=0$, and with
$p_{z}\equiv k$ one obtains the result in Eq. \eqref{eq:4.9}. Thus, the conserved axial momentum is not associated with $\dot z$ alone: the screw geometry locks the axial and azimuthal coordinate velocities
into the invariant combination $\dot z+\beta\dot\phi$. The conserved quantity $k$ therefore represents the momentum conjugate to the cyclic coordinate $z$, while the combination $\dot z+\beta\dot\phi$ is the
corresponding intrinsic screw-direction velocity. The screw dislocation consequently does not destroy axial momentum conservation; instead, it geometrically redistributes the conserved axial motion between the $z$ and $\phi$ coordinates. This coupling is precisely what produces the shifted combination $\mathcal{J}=\ell-\beta k$ in the radial sector and thereby transfers the global information carried by the screw geometry into the effective one-dimensional radial dynamics.

The radial equation of motion follows directly from the reduced Hamiltonian
and can be written as
\begin{equation}
\ddot r
=
\frac{\mathcal{J}^{2}}{\mu^{2}r^{3}}
-
\frac{a^{2}}{\mu^{2}}\,r.
\label{eq:4.41}
\end{equation}
The two terms have complementary dynamical origins. The first is a
short-distance repulsion generated by the conserved screw-shifted momentum
$\mathcal{J}=\ell-\beta k$, whereas the second is a linear restoring force generated
by the ambient magnetic field through $a=q\mathcal{B}_{0}/2$. Their
competition fixes the finite equilibrium radius
$r_{\mathrm{c}}^{2}=|\mathcal{J}|/|a|$ whenever $a\neq0$ and $\mathcal{J}\neq0$. Importantly,
no independent term proportional to $\beta$ appears in the radial force.
This is a consequence of the exact form of the regular spiral-dislocation
metric: away from the defect core, the geometry is locally flat, while the
nontrivial screw structure survives globally through the conserved
combination $\ell-\beta k$. The electromagnetic contribution is likewise
entirely determined by the pullback of the uniform ambient field and
introduces no additional phenomenological $\beta$-dependent force. Thus,
the screw parameter influences the radial motion through the conserved
momentum sector rather than through a separate local force term.

The role of the magnetic field becomes especially transparent in the
field-free limit. For $a=0$, the radial effective potential reduces to
\begin{equation}
\mathcal{V}_{\mathrm{eff}}(r)
=
\frac{\mathcal{J}^{2}}{2\mu r^{2}},
\label{eq:4.43}
\end{equation}
and the large-radius restoring term disappears. The radial motion is then
asymptotically unconfined, with
\begin{equation}
\mathcal{V}_{\mathrm{eff}}(r)
\longrightarrow
0,
\qquad
r\rightarrow\infty.
\end{equation}
Nevertheless, for $\mathcal{J}\neq0$ the excluded defect axis remains dynamically
inaccessible because of the inverse-square barrier. Only in the simultaneous
limit $a=0$ and $\mathcal{J}=0$ does the reduced radial potential vanish identically,
so that the radial degree of freedom becomes free on the regular domain
$r>0$. This limiting case cleanly separates the two mechanisms responsible
for the nontrivial radial dynamics: the screw-induced momentum barrier and
the magnetic confinement.

Combining the conserved momenta with the pulled-back electromagnetic
connection gives 
\begin{equation}
\mathcal{H}_{r}=\frac{p_{r}^{2}}{2\mu}+\frac{(\mathcal{J}-ar^{2})^{2}}{2\mu r^{2}}.\label{eq:4.46}
\end{equation}
Equation~\eqref{eq:4.46} is the exact classical radial reduction of the
minimally coupled charged-particle dynamics on the regular
spiral-dislocation geometry. All dependence on the global defect
structure is compressed into the conserved combination
$\mathcal{J}=\ell-\beta k$, while the electromagnetic scale is fixed by the ambient
field through $a=q\mathcal{B}_{0}/2$. The resulting one-dimensional problem
therefore contains no additional curvature potential and no independent
local $\beta$-dependent force. Instead, the global screw holonomy shifts
the effective angular momentum, while the uniform ambient magnetic field
supplies the large-radius restoring scale. Their combined contribution produces the inverse-square short-distance barrier, the finite equilibrium radius, and, for nonzero magnetic field, the asymptotic quadratic confinement established above. In this sense, the nontrivial radial dynamics is generated not by local curvature of the regular exterior, but by the combined action of the global geometric
momentum shift and the electromagnetic connection inherited from the ambient uniform field.

\begin{figure*}[t]
\centering
\includegraphics[width=1\linewidth]{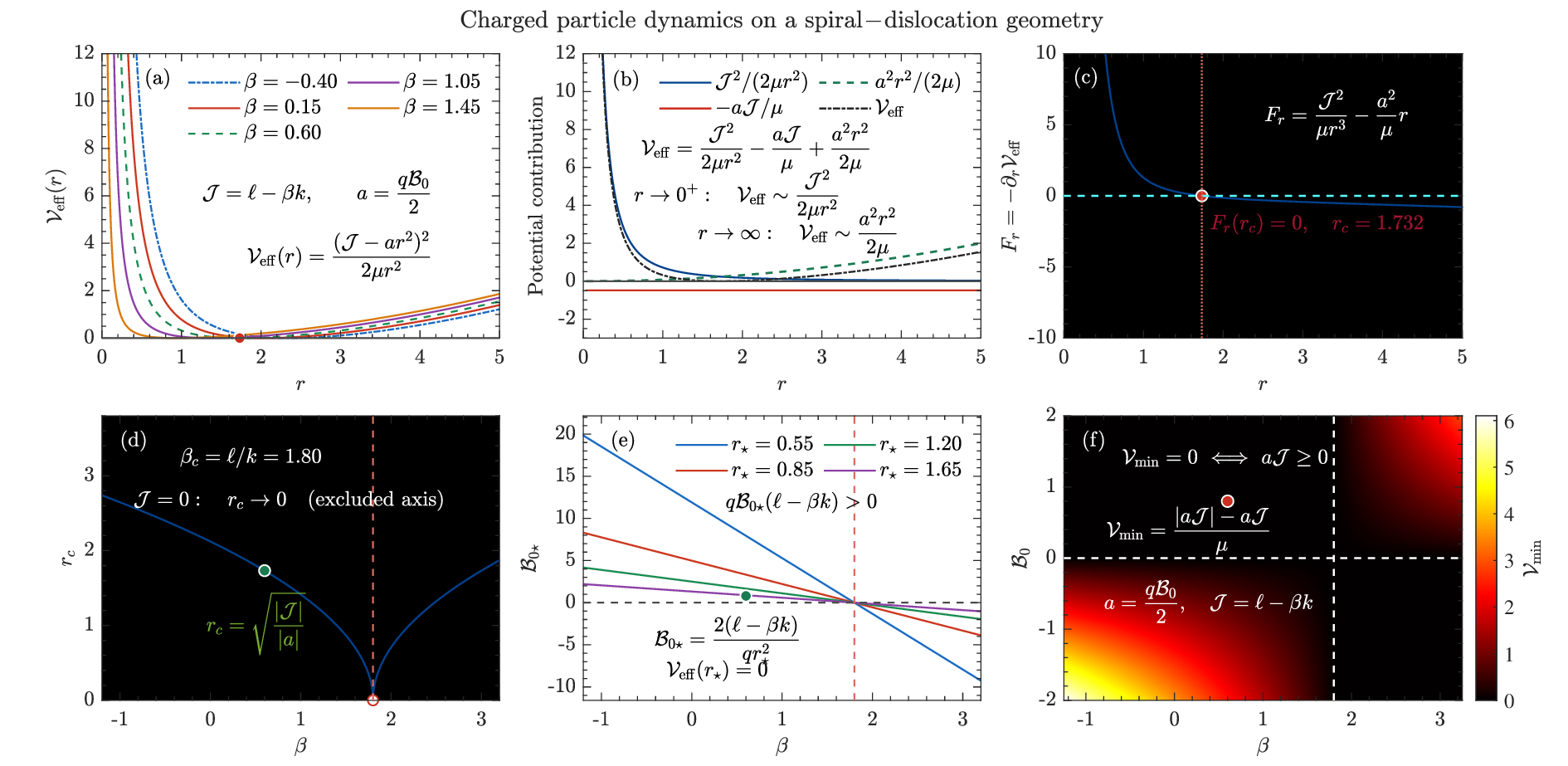}
\caption{\fontsize{7.6}{8.6}\selectfont
Screw--magnetic confinement and zero-potential compensation on the regular spiral-dislocation exterior. The reduced radial dynamics is governed by Eq.~\eqref{eq:4.46}. Unless stated otherwise, the reference parameters are $\mu=1$, $q=1$, $\ell=1.80$, $k=1.00$, $\beta=0.60$, and $\mathcal{B}_{0}=0.80$, yielding $\mathcal{J}=1.20$ and $a=0.40$. (a) Effective radial potential for
$\beta=-0.40,\,0.15,\,0.60,\,1.05,$ and $1.45$ at fixed
$\ell$, $k$, and $\mathcal{B}_{0}$. The finite stationary radius is
$r_{\mathrm{c}}=\sqrt{|\mathcal{J}|/|a|}$; for the reference parameters,
$a\mathcal{J}>0$, so that $r_{\mathrm{c}}$ coincides with the
zero-potential radius $r_{\star}=\sqrt{\mathcal{J}/a}$. (b) Exact decomposition of Eq.~\eqref{eq:4.19} into the inverse-square screw-shifted centrifugal contribution $\mathcal{J}^{2}/(2\mu r^{2})$, the $r$-independent magnetic cross term $-a\mathcal{J}/\mu$, and the quadratic magnetic contribution
$a^{2}r^{2}/(2\mu)$. (c) Radial force obtained from Eq.~\eqref{eq:4.41}, with the unique finite force-balance point $r=r_{\mathrm{c}}$ for $a\neq0$ and $\mathcal{J}\neq0$. (d) Stationary radius as a function of the screw parameter, \(r_{\mathrm{c}}=\sqrt{\frac{|\ell-\beta k|}{|q\mathcal{B}_{0}/2|}}\), with the critical value $\beta_{\mathrm{c}}=\ell/k$ at which $\mathcal{J}=0$. At this value the finite-radius branch approaches $r=0$; since the defect axis is excluded from the regular exterior, this limiting point is not a regular finite-radius equilibrium. (e) Ambient magnetic field required to impose
$\mathcal{V}_{\mathrm{eff}}(r_{\star})=0$ at a prescribed radius, \(\mathcal{B}_{0\star}
=\frac{2(\ell-\beta k)}{q r_{\star}^{2}}\), for $r_{\star}=0.55,\,0.85,\,1.20,$ and $1.65$. The reference point is $(\beta,\mathcal{B}_{0})=(0.60,0.80)$. A finite zero of $\mathcal{V}_{\mathrm{eff}}$ exists only in the sign-compatible sector $q\mathcal{B}_{0}(\ell-\beta k)>0$, in accordance with
Eq.~\eqref{eq:4.96}. (f) Parameter-space map of the global minimum
\(\mathcal{V}_{\min}=\frac{|a\mathcal{J}|-a\mathcal{J}}{\mu}\) from Eq.~\eqref{eq:4.33} in the $(\beta,\mathcal{B}_{0})$ plane. The lines $\ell-\beta k=0$ and $\mathcal{B}_{0}=0$ separate regions with
opposite signs of $a\mathcal{J}$. The minimum is zero for $a\mathcal{J}\geq0$, with a finite zero-potential radius for $a\mathcal{J}>0$, whereas $a\mathcal{J}<0$ gives a strictly positive minimum. Throughout, the radial coordinate is restricted to the regular domain $r>0$, with the defect axis $r=0$ excluded.
}
\label{fig:confinement}
\end{figure*}

\subsection{\mdseries{Zero-Energy Threshold and Screw--Magnetic Compensation}}
\label{sec:zero_energy_threshold}

\setlength{\parindent}{0pt}

The radial potential derived in Eq.~\eqref{eq:4.19} admits a particularly
transparent interpretation when written in its unexpanded form,
\begin{equation}
\mathcal{V}_{\mathrm{eff}}(r)
=
\frac{(\mathcal{J}-ar^{2})^{2}}{2\mu r^{2}}.
\label{eq:4.79}
\end{equation}
Because $\mu>0$ and $r>0$ on the regular exterior, Eq.~\eqref{eq:4.79}
implies the strict non-negativity condition
\begin{equation}
\mathcal{V}_{\mathrm{eff}}(r)\geq0.
\label{eq:4.80}
\end{equation}
Consequently, the lower bound of the reduced radial potential is reached
at zero if and only if the numerator of Eq.~\eqref{eq:4.79} vanishes.
The corresponding condition is
\begin{equation}
\mathcal{J}-ar_{\star}^{2}=0,
\label{eq:4.81}
\end{equation}
and therefore
\begin{equation}
r_{\star}^{2}
=
\frac{\mathcal{J}}{a}
=
\frac{2(\ell-\beta k)}
{q\mathcal{B}_{0}}.
\label{eq:4.82}
\end{equation}
A finite regular solution requires $r_{\star}^{2}>0$, which gives the
existence condition
\begin{equation}
a\mathcal{J}>0,
\qquad\text{or equivalently}\qquad
q\mathcal{B}_{0}(\ell-\beta k)>0.
\label{eq:4.83}
\end{equation}
Thus, exact cancellation is possible only when the magnetic coupling and
the screw-shifted conserved angular momentum have the same sign. When
this condition is satisfied, the finite radius
\begin{equation}
r_{\star}
=
\sqrt{\frac{2(\ell-\beta k)}
{q\mathcal{B}_{0}}}
=
\sqrt{\frac{|\mathcal{J}|}{|a|}}
\label{eq:4.84}
\end{equation}
is a regular point of the configuration space at which
\begin{equation}
\mathcal{V}_{\mathrm{eff}}(r_{\star})=0.
\label{eq:4.85}
\end{equation}

This zero has a direct dynamical meaning. According to the radial Hamiltonian given in Eq. \eqref{eq:4.46}, the condition $\mathcal{V}_{\mathrm{eff}}(r_{\star})=0$ coincides with the
absolute minimum of the reduced radial potential. At the corresponding threshold energy, the radial kinetic term must also vanish, $p_{r}=0$, so that the radial motion is stationary at $r=r_{\star}$.
The zero is therefore not an ordinary crossing of the potential through
the energy level; rather, it represents a degenerate radial turning point
and, equivalently, the finite-radius equilibrium configuration identified
by the minimum analysis above. The azimuthal and axial motions may remain
nontrivial, since the conserved quantities $\ell$ and $k$ continue to govern the cyclic degrees of freedom. The compensation condition can also be expressed directly as a balance between the screw-shifted canonical angular momentum and the magnetic field contribution,
\begin{equation}
\ell-\beta k
=
\frac{q\mathcal{B}_{0}}{2}\,r_{\star}^{2}.
\label{eq:4.87}
\end{equation}
This form makes the geometric content of the threshold especially
transparent. The screw dislocation does not generate an additional local
radial force proportional to $\beta$. Instead, its entire influence on
the reduced radial sector is encoded in the globally shifted conserved
quantity $\ell-\beta k$. The ambient magnetic field then supplies the
radially growing contribution against which this shifted momentum can be
balanced. The zero of the radial potential therefore occurs precisely
when these two contributions cancel at a finite radius. For fixed nonzero $\mathcal{J}$, the orientation of the magnetic coupling is
essential. If
\begin{equation}
a\mathcal{J}<0,
\label{eq:4.88}
\end{equation}
then $\mathcal{J}$ and $ar^{2}$ have opposite signs for every $r>0$, and the
condition in Eq.~\eqref{eq:4.81} cannot be satisfied. The effective
potential consequently has no finite zero and remains strictly positive,
with its minimum given by Eq.~\eqref{eq:4.33}. Hence, reversing the sign
of $q\mathcal{B}_{0}$ at fixed $\ell$, $k$, and $\beta$ can switch the
system between a finite-radius zero-potential branch and a strictly
positive radial minimum. The threshold is therefore intrinsically
orientation-selective: it depends on the relative sign of the ambient
magnetic coupling and the screw-shifted momentum, rather than on their
magnitudes alone. The finite-radius branch possesses a distinguished limiting point. From
Eq.~\eqref{eq:4.84},
\begin{equation}
r_{\star}^{2}
=
\frac{|\mathcal{J}|}{|a|}
\longrightarrow0
\qquad
\text{as}
\qquad
\mathcal{J}=\ell-\beta k\longrightarrow0,
\label{eq:4.89}
\end{equation}
for fixed nonzero $a$. Thus the zero-potential equilibrium is continuously
driven toward the defect axis as the screw-shifted momentum is tuned to
zero. However, the axis $r=0$ is excluded from the regular spiral-
dislocation exterior. Consequently, the limiting condition \(\ell-\beta k=0\) does not define an additional regular equilibrium. Rather, it marks the boundary of the finite-radius zero-potential branch, where that branch terminates at the excluded defect locus. For $k\neq0$, the same condition
takes the form \(\frac{\ell}{k}=\beta\) showing explicitly that the endpoint is reached when the azimuthal canonical momentum is exactly compensated by the screw-induced axial momentum shift. The limiting behavior can also be seen directly from the effective
potential. When $\mathcal{J}=0$,
\begin{equation}
\mathcal{V}_{\mathrm{eff}}(r)
=
\frac{a^{2}r^{2}}{2\mu},
\qquad
r>0,
\label{eq:4.92}
\end{equation}
so that the potential approaches zero only in the excluded limit
$r\rightarrow0^{+}$. For nonzero $a$, the origin is therefore not a
finite regular zero of the radial problem. The disappearance of the
inverse-square sector at $\mathcal{J}=0$ does not create a new equilibrium in the
regular exterior; instead, it removes the centrifugal barrier while
leaving the magnetic quadratic confinement intact. This limiting case
provides a sharp distinction between a genuine finite-radius
screw--magnetic compensation point and the singular boundary of the
configuration space. It is also useful to distinguish the zero of the reduced radial potential
from the total conserved energy of the particle. After eliminating the
cyclic coordinates, the full Hamiltonian may be written as
\begin{equation}
\mathcal{H}
=
\frac{k^{2}}{2\mu}
+
\mathcal{H}_{r},
\qquad
\mathcal{H}_{r}
=
\frac{p_{r}^{2}}{2\mu}
+
\mathcal{V}_{\mathrm{eff}}(r).
\label{eq:4.93}
\end{equation}
Accordingly, the condition
$\mathcal{V}_{\mathrm{eff}}(r_{\star})=0$ denotes a
\emph{zero of the reduced radial potential}, or equivalently the
zero-threshold configuration of the radial sector. It does not imply
that the total particle energy vanishes when $k\neq0$. At the equilibrium
point,
\begin{equation}
p_{r}=0,
\qquad
\mathcal{V}_{\mathrm{eff}}(r_{\star})=0,
\qquad
\mathcal{H}_{\star}
=
\frac{k^{2}}{2\mu},
\label{eq:4.94}
\end{equation}
for the Hamiltonian convention used here. This separation is important:
the threshold concerns the radial degree of freedom after the conserved
axial motion has been isolated, whereas the full energy retains the
independent contribution associated with the cyclic coordinate $z$. The magnetic field required to realize a zero-potential configuration at a
prescribed regular radius can be obtained directly from
Eq.~\eqref{eq:4.87},
\begin{equation}
\mathcal{B}_{0}
=
\frac{2(\ell-\beta k)}
{q r_{\star}^{2}},
\label{eq:4.95}
\end{equation}
with the sign constraint \(q\mathcal{B}_{0}(\ell-\beta k)>0\). Hence the ambient field does not merely determine the strength of radial confinement; its magnitude and orientation jointly select the location
and existence of the zero-threshold configuration. In particular, for fixed $\ell$, $k$, $\beta$, and $r_{\star}$, the required field is set uniquely by the screw-shifted momentum $\ell-\beta k$. The significance of this threshold follows from the geometric construction of the problem. On the regular exterior, the Levi-Civita curvature vanishes identically, so there is no smooth curvature-induced
potential capable of producing the compensation condition. The nontrivial geometric information instead survives globally through the screw identification and enters the conserved dynamics through
$\mathcal{J}=\ell-\beta k$. Independently, the electromagnetic sector is inherited
from the ambient uniform field by the pullback construction, fixing
$a=q\mathcal{B}_{0}/2$ without introducing any additional local
$\beta$-dependent gauge interaction. The zero of $\mathcal{V}_{\mathrm{eff}}$ is therefore generated by the exact competition of these two structures: \(\ell-\beta k=\frac{q\mathcal{B}_{0}}{2}r_{\star}^{2}\). The finite-radius branch exists only for the sign-compatible sector
$q\mathcal{B}_{0}(\ell-\beta k)>0$, while its radius contracts
continuously toward the excluded defect axis as
$\ell-\beta k\rightarrow0$. The resulting threshold is therefore a
direct dynamical manifestation of the global screw structure in an
ambient uniform magnetic field, obtained entirely from minimal coupling
and the intrinsic geometry, without the introduction of any
phenomenological core potential or local curvature force.

Figure~\ref{fig:confinement} summarizes the reduced radial dynamics obtained from Eqs.~\eqref{eq:4.16}--\eqref{eq:4.21} and the zero-potential condition of Sec.~\ref{sec:zero_energy_threshold}. On the regular domain $r>0$, the exterior geometry is locally flat, so no curvature-induced radial force or curvature potential is present. The defect enters through the conserved combination $\mathcal{J}=\ell-\beta k$ of Eq.~\eqref{eq:4.17}, whereas the magnetic scale is $a=q\mathcal{B}_{0}/2$. For $a\neq0$ and $\mathcal{J}\neq0$, Eqs.~\eqref{eq:4.35} and \eqref{eq:4.36} give $\mathcal{V}_{\mathrm{eff}}\to+\infty$ as $r\to0^{+}$ and $r\to\infty$. Together with the positivity of Eq.~\eqref{eq:4.32}, this establishes a unique finite global minimum. Its position depends on $\beta$ only through $\mathcal{J}$. When
$a\mathcal{J}>0$, the minimum coincides with the finite zero of the reduced potential; when $a\mathcal{J}<0$, the minimum remains strictly positive according to Eq.~\eqref{eq:4.33}. Panel~(a) shows the shift of the radial minimum with $\beta$ at fixed $\ell$, $k$, and $\mathcal{B}_{0}$. Panel~(b) separates the inverse-square, constant, and quadratic terms of Eq.~\eqref{eq:4.19}. The first controls
the short-distance behavior, while the last determines the large-radius confinement. The constant term does not contribute to the radial force. Panel~(c) shows the force from Eq.~\eqref{eq:4.41}; its unique zero gives the stationary radius, whose stability follows from Eq.~\eqref{eq:4.32}.

Panel~(d) shows the stationary radius as a function of $\beta$. For
$k\neq0$, $\mathcal{J}=0$ at $\beta_{\mathrm c}=\ell/k$, and the
finite-radius branch approaches the excluded axis $r=0$. This limiting
point is not a regular equilibrium. Panel~(e) shows the field required
to impose a zero of the reduced potential at a prescribed radius,
as given by Eq.~\eqref{eq:4.95}, subject to the sign condition
regarding \(q\mathcal{B}_{0}(\ell-\beta k)\). Panel~(f) displays the corresponding minimum in the
$(\beta,\mathcal{B}_{0})$ plane. The sectors $a\mathcal{J}>0$ and
$a\mathcal{J}<0$ correspond, respectively, to zero and strictly positive
radial minima.

The zero discussed here refers to the reduced radial potential, not to
the total particle energy. From Eqs.~\eqref{eq:4.93} and
\eqref{eq:4.94}, the full Hamiltonian retains the axial contribution
$k^{2}/(2\mu)$. Thus, the finite-radius threshold is a property of the
radial sector, with its location fixed by the screw-shifted momentum and
the magnetic coupling. The limits $a=0$ and $\mathcal{J}=0$ remove,
respectively, the large-radius magnetic confinement and the inverse-square
barrier, as discussed in Eq.~\eqref{eq:4.43} and Eq.~\eqref{eq:4.92}.

\section{Phase-Space Structure, Winding Sectors, and Topology of the Spiral Dislocation}
\label{sec:spiral_phase_space}

\begin{figure*}[t]
\centering
\includegraphics[width=0.7\linewidth]{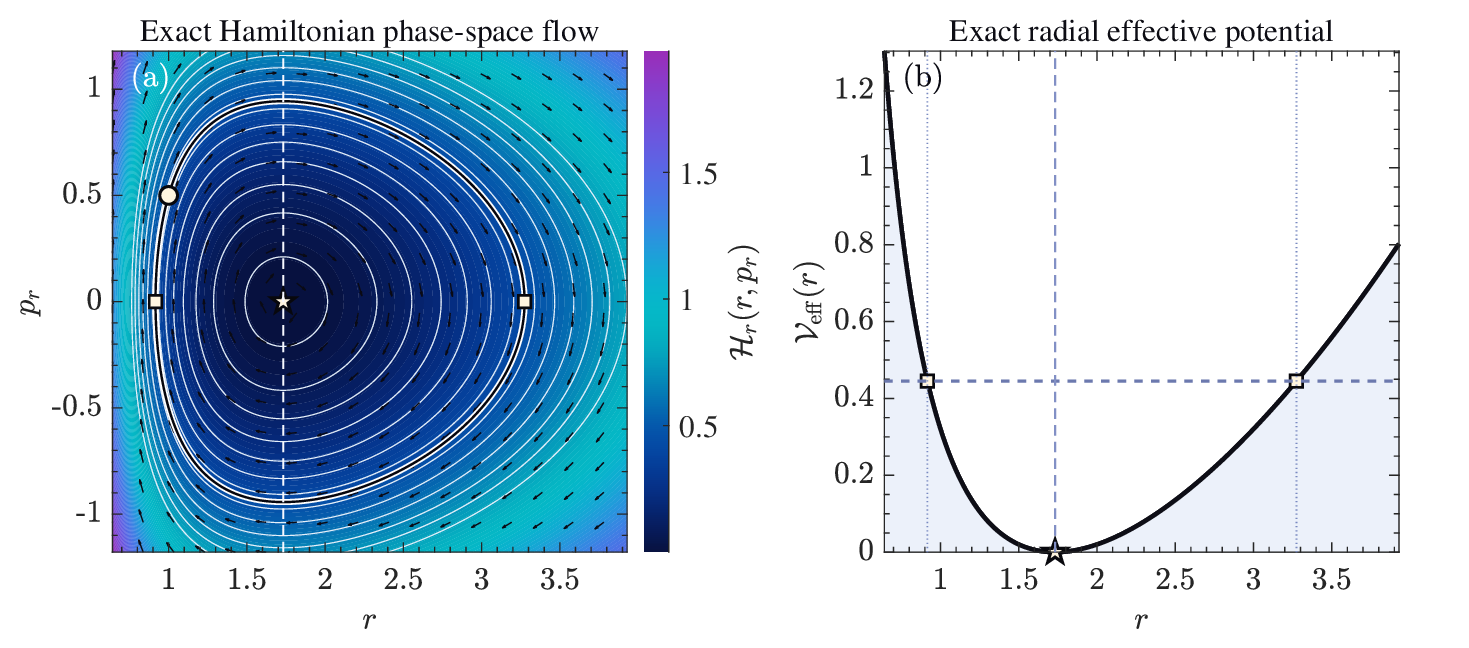}
\caption{\fontsize{7.6}{8.6}\selectfont
Radial Hamiltonian phase-space dynamics and the associated effective radial potential for a charged particle in the regular exterior of a spiral dislocation. The reduced Hamiltonian is
$\mathcal{H}_{r}=p_{r}^{2}/(2\mu)+(\mathcal{J}-ar^{2})^{2}/(2\mu r^{2})$, where $\mathcal{J}=\ell-\beta k$ and $a=q\mathcal{B}_{0}/2$. In panel~(a), the color scale represents the exact Hamiltonian $\mathcal{H}_{r}(r,p_{r})$, while the arrows denote the normalized Hamiltonian vector field generated by
$\dot r=p_{r}/\mu$ and $\dot p_{r}=\mathcal{J}^{2}/(\mu r^{3})-a^{2}r/\mu$. The white contour identifies the conserved energy shell $\mathcal{H}_{r}=H_{0}$, and the superposed curve is the numerical trajectory obtained from the exact radial equations. The point $(r_{\star},0)$ denotes the finite-radius equilibrium, $r_{\star}=\sqrt{|\mathcal{J}|/|a|}$. The marked intersections with $p_{r}=0$ are the radial turning points. Panel~(b) shows the corresponding effective potential
$\mathcal{V}_{\mathrm{eff}}(r)=(\mathcal{J}-ar^{2})^{2}/(2\mu r^{2})$, together with the conserved energy $H_{0}$, the equilibrium point, and the two turning radii determined by
$\mathcal{V}_{\mathrm{eff}}(r_{\pm})=H_{0}$. The parameters are
$\mu=1$, $q=1$, $\mathcal{B}_{0}=0.80$, $\ell=1.80$, $k=1.00$, and $\beta=0.60$, yielding
$\mathcal{J}=1.20$, $a=0.40$, and $r_{\star}=\sqrt{3}\simeq1.73205$. The initial condition is $(r_{0},p_{r0})=(1.00,0.50)$. The radial coordinate is restricted to the regular domain $r>0$, with the defect axis excluded.}\label{fig:phase}
\end{figure*}

\setlength{\parindent}{0pt}

The preceding analysis reduces the charged-particle dynamics to an integrable system whose nontrivial radial structure is governed by the single screw-shifted momentum \(\mathcal{J}\), while the cyclic coordinates $\phi$ and $z$ generate the conserved canonical momenta $\ell$ and $k$. The resulting phase-space organization is therefore more informative than the radial potential alone, since it separates the continuous dynamical invariants from the discrete topological information carried by the screw geometry. In particular, the local flatness established in Sec.~\ref{sec:spiral_geometry} does not imply global triviality. The excluded defect axis and the screw identification generate noncontractible winding sectors that remain dynamically meaningful even though the regular Riemann tensor vanishes identically. The regular spiral-dislocation exterior is consequently locally Euclidean but globally multiply connected. For a closed winding with
\begin{equation}
\Delta\phi=2\pi n,
\qquad
n\in\mathbb{Z},
\label{eq:5.3}
\end{equation}
the screw identification gives
\begin{equation}
\Delta z+\beta\Delta\phi
=
2\pi\beta n
=
n b_{z},
\label{eq:5.5}
\end{equation}
where $b_{z}=2\pi\beta$ is the axial Burgers translation associated with one complete winding. Thus, the Burgers translation introduced in Eq.~\eqref{eq:2.35} is the translational generator associated with one complete winding. The integer $n$ labels the homotopy class, while $\beta$ determines the axial displacement associated with that winding. The screw parameter therefore encodes a global relation between angular winding and axial translation rather than an additional local curvature source.

Because $\phi$ and $z$ are cyclic coordinates in the Hamiltonian derived in Sec.~\ref{sec:spiral_hamiltonian}, the quantities
\begin{equation}
\mathcal H=\mathcal E,
\qquad
p_{\phi}=\ell,
\qquad
p_{z}=k
\label{eq:5.7}
\end{equation}
are conserved. Nevertheless, the radial dynamics does not depend on $\ell$ and $k$ independently. Their geometrically relevant combination is the screw-shifted momentum \(\mathcal{J}=\ell-\beta k\), as established in Eq.~\eqref{eq:4.46}. Once $(\mathcal E,\ell,k)$ are fixed, the full phase-space motion consequently reduces to a one-degree-of-freedom radial flow in the $(r,p_{r})$ plane, together with the reconstruction of the cyclic coordinates. The reduced radial momentum follows directly from the Hamiltonian constraint and can be written as
\begin{equation}
p_{r}
=
\pm
\sqrt{
2\mu
\left[
\mathcal E-\frac{k^{2}}{2\mu}
-\mathcal{V}_{\mathrm{eff}}(r)
\right]
}.
\label{eq:5.10}
\end{equation}
For $a\neq0$ and $\mathcal{J}\neq0$, the inverse-square divergence as $r\rightarrow0^{+}$ and the quadratic growth as $r\rightarrow\infty$, established in Eqs.~\eqref{eq:4.35} and \eqref{eq:4.36}, confine finite-energy radial motion to a finite interval whenever the energy exceeds the radial minimum. The two branches of Eq.~\eqref{eq:5.10} meet at the turning points and form a closed orbit in the reduced radial phase plane. The equilibrium configuration is represented by the fixed point
\begin{equation}
p_{r}=0,
\qquad
r=r_{\star},
\label{eq:5.11}
\end{equation}
with $r_{\star}$ determined by the minimum condition already established in Sec.~\ref{sec:spiral_hamiltonian}. This fixed point belongs to the regular configuration space only when $r_{\star}>0$; the limit $\mathcal{J}\rightarrow0$ instead drives it toward the excluded boundary $r=0$. For a closed configuration-space trajectory, the winding number can equivalently be expressed as
\begin{equation}
\tilde{n}
=
\frac{1}{2\pi}
\oint d\phi
\in\mathbb{Z}.
\label{eq:5.12}
\end{equation}
This integer should be regarded as a topological label rather than as an additional continuous integral of motion. The continuous dynamical invariants are $\mathcal E$, $\ell$, and $k$, whereas $\tilde{n}$ specifies the homotopy class of the associated closed configuration-space trajectory.

The reconstruction of the full orbit further reveals how the global screw structure enters the observable motion. The angular and axial velocities are determined by the conserved quantities and involve the same screw-shifted combination $\mathcal{J}$. Consequently, a radial oscillation is accompanied by correlated azimuthal motion and axial displacement. For a trajectory with winding number $n$, the corresponding screw translation in the lifted embedding space is given by Eq.~\eqref{eq:5.5}. The particle therefore follows a helical family of lifted trajectories even though the regular intrinsic geometry is locally flat. The parameter $\beta$ is consequently observable through the global reconstruction of the orbit rather than through a local curvature force.

The same topological structure also enters the gauge sector through the holonomy of the electromagnetic one-form. For a uniform axial magnetic field, the gauge contribution accumulated over a winding sector is proportional to the enclosed oriented area and takes the form
\begin{equation}
\oint_{\mathcal{C}_{n}}qA
=
\pi q\mathcal{B}_{0}r^{2}n,
\label{eq:5.13}
\end{equation}
with the overall sign fixed by the orientation convention adopted for the loop and the gauge potential. The integer $\tilde{n}$ therefore determines the winding multiplicity of the electromagnetic action contribution, while the magnitude of the contribution is continuously controlled by $q\mathcal{B}_{0}r^{2}$. Topology and electromagnetism thus remain distinct: the former determines the homotopy class of the loop, whereas the latter determines the gauge phase accumulated along that loop.

The phase-space structure is particularly transparent at the zero-energy compensation point discussed in Sec.~\ref{sec:zero_energy_threshold}. The finite-radius zero of the reduced radial potential exists only in the sign-compatible sector
\begin{equation}
q\mathcal{B}_{0}\mathcal{J}>0,
\label{eq:5.14}
\end{equation}
where the magnetic contribution can compensate the screw-shifted momentum at a regular radius. As $\mathcal{J}\rightarrow0$, the equilibrium radius contracts according to the result established previously and approaches the excluded defect axis. Hence, the disappearance of the inverse-square barrier does not provide a continuous dynamical path connecting different winding sectors. Instead, the finite-radius branch terminates at the boundary of the regular configuration space. The compensation mechanism therefore modifies the radial phase-space structure without removing the underlying topological classification.

The global organization of the dynamics can consequently be summarized by distinguishing the continuous and discrete data. The continuous dynamical quantities/invariants $\left(\mathcal E,\ell,k\right)$ determine the local phase-space trajectory and its radial accessibility, whereas $\tilde{n}$ specifies the global homotopy sector of a closed orbit. The screw parameter $\beta$ links these two levels through \(\mathcal{J}=\ell-\beta k\), and is therefore not an additional conserved quantity. Rather, it converts the global translational holonomy of the defect into a shift of the conserved angular-momentum sector that directly controls the radial dynamics.

The resulting phase-space structure is therefore that of an integrable, locally flat system with globally nontrivial topology. The ambient uniform magnetic field then converts the global momentum shift into finite radial confinement and orientation-selective compensation. The spiral dislocation thus provides a particularly transparent realization in which local geometry, global topology, and electromagnetic dynamics remain sharply distinguishable while jointly determining the observable phase-space structure. The nontrivial winding sectors originate from the global configuration space, whereas their dynamical realization is controlled by the conserved screw-shifted momentum and the magnetic confinement mechanism established in the preceding sections.

The phase-space representation in Fig.~\ref{fig:phase} provides the dynamical realization of the exact radial Hamiltonian in Eq.~\eqref{eq:4.46}. Hamilton's equations give
\begin{equation}
\dot r=\frac{p_r}{\mu},
\qquad
\dot p_r=
\frac{\mathcal{J}^{2}}{\mu r^{3}}
-\frac{a^{2}}{\mu}r.
\end{equation}
Hence,
\begin{equation}
\dot p_r\sim
\frac{\mathcal{J}^{2}}{\mu r^{3}}>0
\quad (r\rightarrow0^{+}),
\qquad
\dot p_r\sim
-\frac{a^{2}}{\mu}r<0
\quad (r\rightarrow\infty),
\end{equation}
for $\mathcal{J}\neq0$ and $a\neq0$. The radial flow is therefore directed away from the excluded defect axis at sufficiently small $r$ and toward smaller radii at sufficiently large $r$, yielding bounded finite-energy motion between two positive turning points. The fixed point follows from $\dot p_r=0$ and is precisely the equilibrium radius given in Eq.~\eqref{eq:4.34}. Its stability follows from the positive curvature established in Eq.~\eqref{eq:4.32}. Thus, the closed orbit surrounding the fixed point in panel~(a) represents the exact bounded radial motion, while its intersections with $p_r=0$ are the turning points determined by Eq.~\eqref{eq:4.22}. For the parameters of Fig.~\ref{fig:phase}, $a\mathcal{J}>0$, so the equilibrium coincides with the finite zero of $\mathcal{V}_{\mathrm{eff}}$ described by Eqs.~\eqref{eq:4.81}--\eqref{eq:4.85}. Panel~(b) displays the same radial structure through the effective potential: the inverse-square divergence at $r\rightarrow0^{+}$ and the quadratic growth at $r\rightarrow\infty$ delimit the accessible radial region, while the conserved level $H_{0}$ determines the two turning radii. The numerical trajectory in panel~(a) therefore lies on the corresponding constant-energy shell and provides a direct phase-space representation of the analytic radial dynamics. No local curvature force is involved in this confinement. The regular spiral-dislocation exterior is locally flat, while the defect dependence of the radial sector is entirely contained in the conserved screw-shifted momentum $\mathcal{J}=\ell-\beta k$. The magnetic field enters independently through $a=q\mathcal{B}_{0}/2$ and supplies the large-radius restoring term. The phase portrait thus separates the global geometric contribution from the electromagnetic confinement mechanism: the screw identification modifies the conserved momentum sector, whereas the ambient magnetic field determines the radial confinement scale. No additional core potential or curvature-induced force is required.

\begin{figure*}[t]
\centering
\includegraphics[width=1\textwidth]{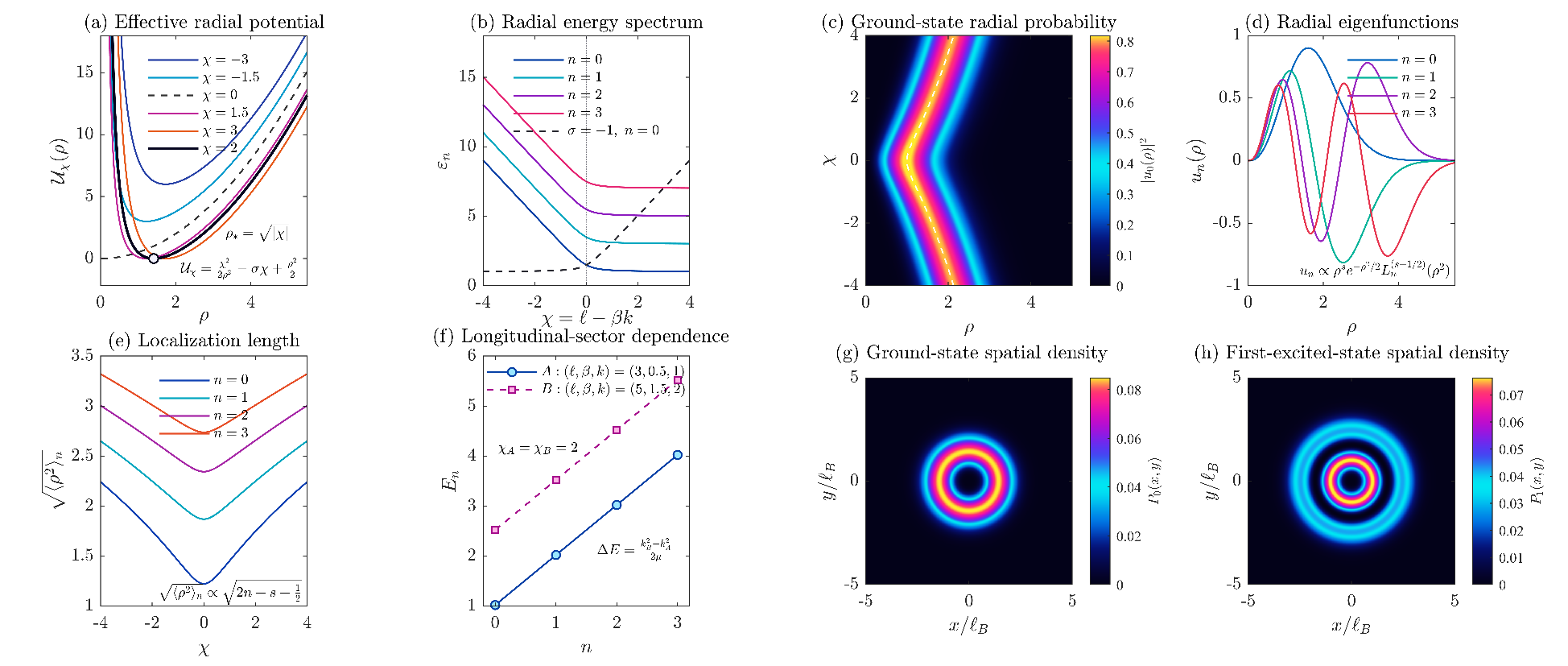}
\caption{\fontsize{8}{9}\selectfont Quantum mapping and spectral structure of a charged particle on the spiral-dislocation geometry in the unit convention $\hbar=1$, with $\sigma=\operatorname{sgn}(q\mathcal B_0)=+1$ and $\mu=1$. (a) Dimensionless effective potential
$\mathcal U_\chi(\rho)=\chi^2/(2\rho^2)-\sigma\chi+\rho^2/2$
for $\chi=-3,-1.5,0,1.5,3$; the $\chi=2$ curve is highlighted and its minimum at $\rho_*=\sqrt{|\chi|}$ is indicated. The horizontal markers on the highlighted potential correspond to the exact levels $n=0,\ldots,3$. (b) Exact dimensionless spectral flow
$\varepsilon_n(\chi)=2n+1+\frac12\sqrt{1+4\chi^2}-\sigma\chi$
for $n=0,\ldots,3$ and $\sigma=+1$, together with the dashed $n=0$ branch for $\sigma=-1$.
(c) Ground-state radial probability density $|u_0(\rho;\chi)|^2$ represented as a continuous map in the $(\rho,\chi)$ plane; the white dashed curve gives the analytical maximum
$\rho_{\rm peak}(\chi)=\left[\frac12(1+\sqrt{1+4\chi^2})\right]^{1/2}$.
(d) Normalized radial eigenfunctions $u_n(\rho)$ for $n=0,\ldots,3$ at $\chi=2$.
(e) Exact dimensionless root-mean-square localization length
$\sqrt{\langle\rho^2\rangle_n}=\sqrt{2n+s+\frac12}$
for $n=0,\ldots,3$, where $s=\frac12(1+\sqrt{1+4\chi^2})$.
(f) Physical energies $E_n$ versus the radial quantum number $n$ for the two parameter sets
$A:(\ell,\beta,k)=(3,0.5,1)$ and $B:(\ell,\beta,k)=(5,1.5,2)$. For the plotted curves, the code evaluates the radial contribution with the common value $\chi_{\rm common}=\chi_A=2.5$, while the distinct longitudinal terms $k^2/(2\mu)$ produce the vertical separation. (g) Ground-state transverse probability density $P_0(x,y)=|u_0(\rho)|^2/(2\pi\rho)$ and (h) first-excited transverse probability density $P_1(x,y)=|u_1(\rho)|^2/(2\pi\rho)$, both evaluated at $\chi=2$ with $\rho=\sqrt{x^2+y^2}/\ell_B$ and $\ell_B=\sqrt{2/|q\mathcal B_0|}$.} \label{fig:QR}
\end{figure*}

\section{Quantum Mapping and Energy Spectrum}
\label{sec:spiral_quantum_radial}

\setlength{\parindent}{0pt}

The classical reduction developed in Sec.~\ref{sec:spiral_hamiltonian} leaves the radial pair $(r,p_r)$ as the only nontrivial dynamical degree of freedom after fixing the conserved axial and azimuthal momenta $k$ and $\ell$. The global screw structure consequently enters the reduced radial dynamics
through the single invariant combination \(\mathcal{J}\equiv\ell-\beta k\). Thus, although $k$ and $\ell$ remain separately conserved quantities of the full system, their appearance in the reduced radial interaction is entirely through $\mathcal{J}$. The longitudinal kinetic contribution $k^2/(2\mu)$
remains separate and is not absorbed into this invariant. The reduced phase space can be
quantized directly on \(L^2((0,\infty),dr)\) with
\begin{equation}
 p_r\longrightarrow-i\frac{d}{dr},\qquad p_r^2\longrightarrow-\frac{d^2}{dr^2}.\label{QM}   
\end{equation}
Accordingly, $u(r)$ denotes the reduced one-dimensional radial amplitude and
is normalized with respect to the measure $dr$. This construction is distinct
from a direct quantization of the unreduced three-dimensional system through
the Laplace--Beltrami operator. In the latter formulation the radial
differential operator and measure follow from the original spatial metric,
whereas here the classical cyclic variables have first been eliminated and
the resulting canonical pair $(r,p_r)$ is subsequently quantized. The
spectrum obtained below therefore refers specifically to this reduced
quantization prescription.

According to the reduced Hamiltonian obtained in Sec.~\ref{sec:spiral_hamiltonian}, the corresponding quantum eigenvalue equation is
\begin{equation}
\left[
-\frac{1}{2\mu}\frac{d^2}{dr^2}
+\frac{k^2}{2\mu}
+\frac{(\mathcal{J}-ar^2)^2}{2\mu r^2}
\right]u(r)
=
\mathcal E\,u(r).
\label{eq:7.4}
\end{equation}
The radial interaction follows from the decomposition
\begin{equation}
\frac{(\mathcal{J}-ar^2)^2}{r^2}
=
\frac{\mathcal{J}^2}{r^2}
-2a\mathcal{J}
+a^2r^2,
\label{eq:7.5}
\end{equation}
so that, using $2a=q\mathcal B_0$, Eq.~(\ref{eq:7.4}) becomes
\begin{equation}
u''(r)
+
\left[
2\mu\mathcal E-k^2
+q\mathcal B_0\mathcal{J}
-\frac{\mathcal{J}^2}{r^2}
-a^2r^2
\right]u(r)
=0.
\label{eq:7.6}
\end{equation}
The three radial contributions are thereby separated into an inverse-square term, a field-dependent constant shift, and a quadratic confining term. For $q\mathcal B_0\neq0$, introduce the positive magnetic parameter
\begin{equation}
\widetilde a\equiv|a|
=
\frac{|q\mathcal B_0|}{2},
\qquad
\widetilde{\mathcal E}
\equiv
2\mu\mathcal E-k^2+q\mathcal B_0\mathcal{J}.
\label{eq:7.7}
\end{equation}
Since $a^2=\widetilde a^{\,2}$, the radial equation takes the canonical
singular-oscillator form \cite{Sakurai,Gottfried}
\begin{equation}
u''(r)
+
\left[
\widetilde{\mathcal E}
-\frac{\mathcal{J}^2}{r^2}
-\widetilde a^{\,2}r^2
\right]u(r)
=0.
\label{eq:7.8}
\end{equation}
The nonzero magnetic field thus introduces the intrinsic radial scale $\widetilde a^{-1/2}$, while the screw geometry enters the singular part of the radial operator through $\mathcal J$. To characterize the singular endpoint at $r=0$, write
\begin{equation}
L(L+1)=\mathcal{J}^2,
\qquad
L=
-\frac12+
\sqrt{\frac14+\mathcal{J}^2},
\qquad
L\geq0.
\label{eq:7.9}
\end{equation}
The local equation near the origin is
\begin{equation}
u''(r)
-
\frac{L(L+1)}{r^2}u(r)
\simeq0,
\qquad
r\rightarrow0^+,
\label{eq:7.10}
\end{equation}
with the two independent behaviors
\begin{equation}
u(r)
\sim
\mathcal{C}_1\,r^{L+1}
+
\mathcal{C}_2\,r^{-L}.
\label{eq:7.11}
\end{equation}
The endpoint is singular, and its treatment is part of the definition of the
quantum Hamiltonian. In the range
\begin{equation}
0\leq L<\frac12
\qquad\Longleftrightarrow\qquad
\mathcal J^2<\frac34,
\label{eq:7.12}
\end{equation}
both local branches are square-integrable near the origin. Hence
square-integrability alone does not select a unique self-adjoint
realization of the inverse-square operator. We choose the Friedrichs
realization, which eliminates the less regular branch and imposes
\begin{equation}
\mathcal{C}_2=0,
\qquad
u(r)\sim r^{L+1},
\qquad
r\rightarrow0^+.
\label{eq:7.13}
\end{equation}
For $L\geq1/2$, the $r^{-L}$ branch is not square-integrable at the origin,
and the same Friedrichs behavior is selected by the endpoint analysis. All
spectral results below refer to this specified self-adjoint realization.

At large radius, the quadratic term in Eq.~(\ref{eq:7.8}) dominates, giving
\begin{equation}
u''(r)
-
\widetilde a^{\,2}r^2u(r)
\simeq0,
\qquad
r\rightarrow\infty.
\label{eq:7.14}
\end{equation}
The normalizable branch has the Gaussian behavior
\begin{equation}
u(r)
\sim
\exp\left(-\frac{\widetilde a r^2}{2}\right).
\label{eq:7.15}
\end{equation}
Combining the Friedrichs behavior at the origin with the decaying asymptotic behavior at infinity suggests the transformation \cite{Sakurai,Gottfried}
\begin{equation}
u(r)
=
r^{L+1}
e^{-\widetilde a r^2/2}
F(z),
\qquad
z\equiv\widetilde a r^2.
\label{eq:7.16}
\end{equation}
Because $\widetilde a>0$, the half-line $r\in(0,\infty)$ is mapped onto $z\in(0,\infty)$. Substitution of Eq.~(\ref{eq:7.16}) into Eq.~(\ref{eq:7.8}) gives
\begin{equation}
zF''(z)
+
\left(
L+\frac32-z
\right)F'(z)
+
\left[
\frac{\widetilde{\mathcal E}}{4\widetilde a}
-\frac{2L+3}{4}
\right]F(z)
=0.
\label{eq:7.17}
\end{equation}
Introducing
\begin{equation}
\eta\equiv L+\frac32,
\qquad
\alpha
\equiv
-\frac{
\widetilde{\mathcal E}
-(2L+3)\widetilde a
}{
4\widetilde a
},
\label{eq:7.18}
\end{equation}
the equation assumes the standard Kummer form \cite{Sakurai,Gottfried}
\begin{equation}
zF''(z)
+
(\eta-z)F'(z)
-
\alpha F(z)
=0.
\label{eq:7.19}
\end{equation}
The solution compatible with the selected behavior at the origin can therefore be written as \cite{Sakurai,Gottfried}
\begin{equation}
u(r)
=
\mathcal N\,
r^{L+1}
e^{-\widetilde a r^2/2}
\,{}_1F_1
\left(
\alpha,\eta,\widetilde a r^2
\right).
\label{eq:7.20}
\end{equation}

For generic $\alpha$, the large-$z$ behavior of ${}_1F_1(\alpha,\eta,z)$
contains an exponentially increasing contribution. Since the prefactor in
Eq.~(\ref{eq:7.20}) contributes only $e^{-z/2}$, that contribution produces
an asymptotic behavior proportional to $e^{z/2}$ and is incompatible with
square-integrability on the half-line. The increasing contribution is
eliminated when the Kummer series terminates, namely when \cite{Sakurai,Gottfried}
\begin{equation}
\alpha=-n,
\qquad
n=0,1,2,\ldots.
\label{eq:7.21}
\end{equation}
At these values the confluent hypergeometric function reduces to an
associated Laguerre polynomial,
\begin{equation}
{}_1F_1(-n,\eta,z)
=
\frac{n!}{(\eta)_n}
L_n^{(\eta-1)}(z),
\label{eq:7.22}
\end{equation}
where $(\eta)_n$ denotes the Pochhammer symbol \cite{Sakurai,Gottfried}. The termination condition
therefore converts the singular oscillator into a discrete exactly solvable radial problem. Using Eq.~(\ref{eq:7.21}) in Eq.~(\ref{eq:7.18}) yields
\begin{equation}
\widetilde{\mathcal E}
=
2\widetilde a
\left(
2n+L+\frac32
\right),
\qquad
n=0,1,2,\ldots.
\label{eq:7.23}
\end{equation}
Restoring the original energy variable gives
\begin{equation}
\mathcal E_n
=
\frac{k^2}{2\mu}
-\frac{q\mathcal B_0}{2\mu}\mathcal J
+
\frac{|q\mathcal B_0|}{2\mu}
\left(
2n+L+\frac32
\right).
\label{eq:7.24}
\end{equation}
Since
\begin{equation}
L+\frac32
=
1+
\frac12\sqrt{1+4\mathcal J^2},
\label{eq:7.25}
\end{equation}
the spectrum in terms of the original conserved quantities becomes
\begin{equation}
\begin{split}
&\mathcal E_n
=
\frac{k^2}{2\mu}
-\frac{q\mathcal B_0}{2\mu}
(\ell-\beta k)
+
\frac{|q\mathcal B_0|}{2\mu}
\left[
2n+1+
\frac12
\sqrt{
1+4(\ell-\beta k)^2
}
\right],\\
& n=0,1,2,\ldots.
\label{eq:7.26}    
\end{split}
\end{equation}
This is the discrete spectrum of the reduced radial Hamiltonian for $q\mathcal B_0\neq0$ in the Friedrichs realization. The screw parameter $\beta$ does not occur independently in the radial eigenvalue problem; its effect is encoded through the invariant $\ell-\beta k$. The longitudinal
kinetic contribution $k^2/(2\mu)$ remains independent of this reduction. It is useful to introduce
\begin{equation}
s\equiv L+1
=
\frac12
\left(
1+\sqrt{1+4\mathcal J^2}
\right),
\label{eq:7.27}
\end{equation}
for which
\begin{equation}
L=s-1,
\qquad
\eta-1=L+\frac12=s-\frac12.
\label{eq:7.28}
\end{equation}
The normalized eigenfunctions can then be written in the compact form
\begin{equation}
u_n(r)
=
\mathcal N_n\,
r^s
e^{-\widetilde a r^2/2}
L_n^{(s-\frac12)}
\left(
\widetilde a r^2
\right).
\label{eq:7.29}
\end{equation}
The normalization condition is
\begin{equation}
\int_0^\infty
|u_n(r)|^2\,dr
=
1.
\label{eq:7.30}
\end{equation}
With
\begin{equation}
z=\widetilde a r^2,
\qquad
dr=
\frac{dz}{2\sqrt{\widetilde a z}},
\label{eq:7.31}
\end{equation}
the normalization integral reduces to the standard associated-Laguerre orthogonality relation. The resulting normalization constant is \cite{Sakurai,Gottfried}
\begin{equation}
\mathcal N_n
=
\left[
\frac{
2\widetilde a^{\,s+\frac12}n!
}{
\Gamma\left(n+s+\frac12\right)
}
\right]^{1/2}.
\label{eq:7.32}
\end{equation}
Thus the complete normalized radial states are
\begin{equation}
u_n(r)
=
\left[
\frac{
2\widetilde a^{\,s+\frac12}n!
}{
\Gamma\left(n+s+\frac12\right)
}
\right]^{1/2}
r^s
e^{-\widetilde a r^2/2}
L_n^{(s-\frac12)}
\left(
\widetilde a r^2
\right).
\label{eq:7.33}
\end{equation}
The polynomial index is
\begin{equation}
s-\frac12
=
L+\frac12,
\label{eq:7.34}
\end{equation}
whereas the radial power $s=L+1$ is fixed by the selected Friedrichs behavior at the singular endpoint. A dimensionless formulation makes the separation between radial structure, magnetic orientation, and longitudinal energy particularly transparent. Considering
\begin{equation}
\sigma
\equiv
\operatorname{sgn}(q\mathcal B_0),
\quad
\chi
\equiv
\mathcal J
=
\ell-\beta k,
\quad
\ell_B
\equiv
\sqrt{\frac{2}{|q\mathcal B_0|}},
\quad
\rho
\equiv
\frac{r}{\ell_B},
\label{eq:7.35}
\end{equation}
the definition of $\ell_B$ gives the identity
\begin{equation}
\widetilde a\,\ell_B^2=1,
\qquad
\widetilde a r^2=\rho^2.
\label{eq:7.36}
\end{equation}
Hence the dimensional magnetic scale is completely removed from the dimensionless radial coordinate. The dimensionless radial energy is defined by subtracting the independent longitudinal kinetic contribution,
\begin{equation}
\varepsilon_n
\equiv
\frac{2\mu}{|q\mathcal B_0|}
\left(
\mathcal E_n-\frac{k^2}{2\mu}
\right).
\label{eq:7.37}
\end{equation}
Since
\begin{equation}
q\mathcal B_0
=
\sigma|q\mathcal B_0|,
\label{eq:7.38}
\end{equation}
Eq.~(\ref{eq:7.26}) becomes
\begin{equation}
\varepsilon_n
=
-\sigma\chi
+
2n+1+
\frac12\sqrt{1+4\chi^2}.
\label{eq:7.39}
\end{equation}
Thus the entire dimensionless radial spectrum is determined by the pair $(\sigma,\chi)$. Different sets of dimensional parameters $(k,\ell,\beta)$ that produce the same value of $\chi$ lead to the same
dimensionless radial eigenvalue problem for fixed $\sigma$. Under the same scaling, Eq.~(\ref{eq:7.8}) becomes
\begin{equation}
\left[
-\frac{d^2}{d\rho^2}
+
\frac{\chi^2}{\rho^2}
-2\sigma\chi
+\rho^2
\right]
u(\rho)
=
2\varepsilon\,u(\rho).
\label{eq:7.40}
\end{equation}
Equivalently,
\begin{equation}
\left[
-\frac12\frac{d^2}{d\rho^2}
+
\mathcal U_\chi(\rho)
\right]
u(\rho)
=
\varepsilon\,u(\rho),
\label{eq:7.41}
\end{equation}
where
\begin{equation}
\mathcal U_\chi(\rho)
=
\frac{\chi^2}{2\rho^2}
-\sigma\chi
+\frac{\rho^2}{2}.
\label{eq:7.42}
\end{equation}
The singular term is controlled by $\chi^2$, whereas the constant magnetic term is linear in $\chi$ and changes sign under $\chi\rightarrow-\chi$. The quadratic term supplies confinement at large $\rho$. Thus the radial Hamiltonian is a singular oscillator with a constant orientation-dependent energy shift. For $\chi\neq0$, the stationary point of the effective potential satisfies
\begin{equation}
\frac{d\mathcal U_\chi}{d\rho}
=
-\frac{\chi^2}{\rho^3}
+\rho
=
0,
\label{eq:7.43}
\end{equation}
and hence
\begin{equation}
\rho_{\min}^4=\chi^2,
\qquad
\rho_{\min}^2=|\chi|.
\label{eq:7.44}
\end{equation}
The corresponding minimum of the effective potential is
\begin{equation}
\mathcal U_{\chi,\min}
=
|\chi|-\sigma\chi.
\label{eq:7.45}
\end{equation}
This quantity is a property of the classical effective potential appearing in the reduced quantum Hamiltonian and should not be identified with a quantum eigenvalue. The ground-state eigenvalue is instead
\begin{equation}
\varepsilon_0
=
1+
\frac12\sqrt{1+4\chi^2}
-\sigma\chi.
\label{eq:7.46}
\end{equation}
For large $|\chi|$,
\begin{equation}
\frac12\sqrt{1+4\chi^2}
=
|\chi|
+
\frac{1}{8|\chi|}
+
O(|\chi|^{-3}),
\label{eq:7.47}
\end{equation}
and consequently
\begin{equation}
\varepsilon_0
=
\mathcal U_{\chi,\min}
+
1
+
\frac{1}{8|\chi|}
+
O(|\chi|^{-3}).
\label{eq:7.48}
\end{equation}
The quantum ground state therefore approaches the minimum of the effective potential with a finite radial zero-point contribution in the large-$|\chi|$ regime. The radial size can be obtained exactly from the normalized Laguerre states. Using the associated-Laguerre moments gives \cite{Sakurai,Gottfried}
\begin{equation}
\left\langle r^2\right\rangle_n
=
\int_0^\infty
r^2|u_n(r)|^2\,dr
=
\frac{1}{\widetilde a}
\left(
2n+s+\frac12
\right).
\label{eq:7.49}
\end{equation}
Since
\begin{equation}
\widetilde a
=
\frac{|q\mathcal B_0|}{2},
\label{eq:7.50}
\end{equation}
one obtains
\begin{equation}
\left\langle r^2\right\rangle_n
=
\frac{2}{|q\mathcal B_0|}
\left(
2n+s+\frac12
\right).
\label{eq:7.51}
\end{equation}
Therefore,
\begin{equation}
\sqrt{\left\langle r^2\right\rangle_n}
=
\sqrt{\frac{2}{|q\mathcal B_0|}}
\sqrt{
2n+s+\frac12
}.
\label{eq:7.52}
\end{equation}
At fixed $(n,k,\ell,\beta)$, the quantity $s$ is independent of the
magnitude of the magnetic field, and hence
\begin{equation}
\sqrt{\left\langle r^2\right\rangle_n}
\propto
|q\mathcal B_0|^{-1/2}.
\label{eq:7.53}
\end{equation}
The dimensional localization scale is consequently set by the effective magnetic length $\ell_B$, while the dimensionless radial profile is determined by $|\chi|$ and the radial quantum number $n$. The exact spectrum also exhibits a field-controlled, equally spaced radial
ladder. From Eq.~(\ref{eq:7.39}),
\begin{equation}
\varepsilon_{n+1}-\varepsilon_n=2,
\label{eq:7.54}
\end{equation}
and therefore
\begin{equation}
\mathcal E_{n+1}-\mathcal E_n
=
\frac{|q\mathcal B_0|}{\mu}.
\label{eq:7.55}
\end{equation}
The spacing is independent of $\chi$, $\beta$, and $\ell$ at fixed $|q\mathcal B_0|$. These quantities instead determine the spectral offset through the singular contribution and the orientation-dependent term. The spatial and spectral dependence on the screw invariant separate exactly.
Since
\begin{equation}
s(\chi)
=
\frac12
\left(
1+\sqrt{1+4\chi^2}
\right),
\label{eq:7.56}
\end{equation}
one has
\begin{equation}
s(-\chi)=s(\chi).
\label{eq:7.57}
\end{equation}
Consequently, for fixed $n$ and fixed magnetic-field magnitude, the normalized radial densities satisfy
\begin{equation}
|u_{n,\chi}(\rho)|^2
=
|u_{n,-\chi}(\rho)|^2.
\label{eq:7.58}
\end{equation}
The dimensionless energies, in contrast, obey
\begin{equation}
\varepsilon_n(-\chi)
=
+\sigma\chi
+
2n+1+
\frac12\sqrt{1+4\chi^2},
\label{eq:7.59}
\end{equation}
and therefore
\begin{equation}
\varepsilon_n(-\chi)-\varepsilon_n(\chi)
=
2\sigma\chi.
\label{eq:7.60}
\end{equation}
The dependence of the radial eigenfunctions on the screw invariant is thus even in $\chi$, whereas the field-orientation contribution to the energy is odd in $\chi$. This separation can be displayed directly as
\begin{equation}
\varepsilon_n=2n+1+
\frac12\sqrt{1+4\chi^2}-\sigma\chi.
\label{eq:7.61}
\end{equation}
The first term determines the radial eigenfunction through $\chi^2$, while the second term shifts the eigenvalue without modifying the radial wavefunction. Equal values of $\chi$ therefore define identical dimensionless radial Hamiltonians for fixed $\sigma$. This equivalence concerns the reduced
dimensionless radial problem and does not imply equality of the complete physical energies for arbitrary underlying parameter sets. Indeed, the physical spectrum may be written as
\begin{equation}
\mathcal E_n
=
\frac{k^2}{2\mu}
+
\frac{|q\mathcal B_0|}{2\mu}
\left[
2n+1+
\frac12\sqrt{1+4\chi^2}
-\sigma\chi
\right].
\label{eq:7.62}
\end{equation}
Hence two parameter sets may have the same $\chi$ and consequently the same dimensionless radial spectrum while carrying different values of the independent longitudinal contribution $k^2/(2\mu)$. The equality of the reduced radial problems must therefore be distinguished from equality of the
full physical energies. The zero-field sector must be treated separately. When $q\mathcal B_0=0$, one has $a=0$, and the magnetic oscillator scale $\widetilde a^{-1/2}$ is absent. Equation~(\ref{eq:7.4}) then reduces to
\begin{equation}
\left[
-\frac{1}{2\mu}\frac{d^2}{dr^2}
+
\frac{k^2}{2\mu}
+
\frac{\mathcal J^2}{2\mu r^2}
\right]
u(r)
=
\mathcal E\,u(r).
\label{eq:7.63}
\end{equation}
After subtracting the constant longitudinal contribution, the radial operator is
\begin{equation}
H_r^{(0)}
=
-\frac{1}{2\mu}\frac{d^2}{dr^2}
+
\frac{\mathcal J^2}{2\mu r^2}.
\label{eq:7.64}
\end{equation}
This operator contains no intrinsic radial length scale. For the Friedrichs
realization specified above, the nonnegative inverse-square interaction does
not generate a discrete oscillator-like bound-state spectrum; the radial
spectrum is continuous above the longitudinal threshold
\begin{equation}
\mathcal E_{\rm th}
=
\frac{k^2}{2\mu}.
\label{eq:7.65}
\end{equation}
Accordingly, the discrete levels in Eq.~(\ref{eq:7.26}) belong exclusively to
the nonzero-field problem. The limit $q\mathcal B_0\rightarrow0$ must be
taken with care because the effective magnetic length \(\ell_B=\sqrt{\frac{2}{|q\mathcal B_0|}}\) diverges and the discrete oscillator scale disappears.

Figure~\ref{fig:QR} provides a direct numerical representation of the reduced quantum problem and its exact spectral and localization properties. Panel~(a) shows the singular-oscillator potential generated by the invariant $\chi=\ell-\beta k$, including the centrifugal barrier for $\chi\neq0$ and the confining quadratic term. For the highlighted $\chi=2$ sector, the minimum occurs at $\rho_*=\sqrt{|\chi|}$, while the marked energies coincide with the exact levels $n=0,\ldots,3$. Panel~(b) displays the corresponding spectral flow, with the radial levels remaining equally spaced by $\Delta\varepsilon=2$ while the $\chi$-dependent term produces the branch displacement and the reversal under $\sigma\rightarrow-\sigma$. Panel~(c) resolves the ground-state radial probability continuously in the $(\rho,\chi)$ plane; its maximum follows the analytical ridge $\rho_{\rm peak}(\chi)$, demonstrating the displacement of the localized radial profile with the reduced angular invariant. The individual normalized eigenfunctions for $\chi=2$ are shown in panel~(d), where increasing $n$ introduces the expected radial nodes. Panel~(e) quantifies the corresponding localization scale through the root-mean-square radius, showing its dependence on both radial excitation and $|\chi|$. Panel~(f) gives the physical energy levels for the two parameter sets used in the numerical construction. Here the radial contribution is deliberately evaluated at the common value $\chi_{\rm common}=2.5$, whereas the different values of $k$ generate distinct longitudinal offsets $k^2/(2\mu)$. Panels~(g) and (h) reconstruct the associated transverse probability densities for the ground and first-excited states at $\chi=2$; their radial symmetry follows directly from the reduced eigenfunctions.

\section{Summary and discussion}
\label{sec:discussion}

\setlength{\parindent}{0pt}

The results provide a unified description of how a globally nontrivial but
locally Euclidean screw geometry modifies the motion of a charged particle
in a uniform magnetic field. The central point is that the dislocation does
not introduce local curvature: the three-dimensional embedding is Euclidean
away from the defect, while the nontriviality resides in the screw identification
\begin{equation}
\phi\rightarrow\phi+2\pi,
\qquad Z\to Z+2\pi\beta .
\label{eq:disc_identification}
\end{equation}
Accordingly, the geometric information relevant to the particle dynamics is
carried by the globally induced coupling between the axial and azimuthal
momenta rather than by a curvature potential. This distinction is essential
for interpreting both the classical and quantum results.

The pullback of the uniform ambient magnetic field to the defect geometry
provides the corresponding gauge coupling. In the intrinsic coframe,
the magnetic potential retains its azimuthal character, while the screw
one-form \(dz+\beta d\phi\) couples the axial and angular sectors. As a
result, the two cyclic momenta remain conserved and the radial dynamics
depends on them only through
\begin{equation}
\mathcal{J}\equiv\ell-\beta k .
\label{eq:disc_invariant}
\end{equation}
This combination is the natural invariant of the screw geometry: changing
the screw parameter does not generate an independent local radial
interaction, but instead shifts the effective angular momentum carried by
the particle.

It is instructive to place this holonomy explicitly in the context of the published literature on geometric and Berry phases for screw dislocations. The translational holonomy of the defect, $\oint(dZ-dz)=2\pi\beta$, is a purely torsional effect: it enters the particle dynamics not as an external vector potential threading a flux tube, but through the metric itself, exactly as first established for the Landau problem in a screw-dislocation background in \cite{furtado1999landau} and for the associated scattering problem in \cite{furtado2000berry}. In those works, and in the related treatment of a screw dislocation carrying an additional internal magnetic flux \cite{furtado2001scattering}, the geometric holonomy manifests as an Aharonov--Bohm-type phase accumulated by scattering (continuum) states, together with a genuine Berry phase associated with the adiabatic transport of the electron around the defect; in \cite{furtado2001scattering} this geometric contribution can even be tuned to cancel exactly against the phase produced by the internal flux, so that neither scattering nor a Berry phase survives for a specific matching condition. The solid-state analog of the He--McKellar--Wilkens phase obtained for a quasiparticle in a charged screw dislocation immersed in a uniform magnetic field \cite{delimaribeiro2003hmw} likewise identifies the topological content of the combined dislocation-plus-field configuration through an explicit geometric phase in the scattering amplitude.

The present system realizes the same combined screw-and-gauge holonomy, but in a regime where the particle is not scattered but confined: the uniform magnetic field converts the open radial problem into a discrete spectrum, so the holonomy no longer appears as a phase multiplying an asymptotic wave, but as a shift of the quantum numbers themselves. Concretely, the geometric holonomy identified in \cite{furtado1999landau,furtado2000berry} enters here through the same mechanism, namely the shift $\ell\rightarrow \mathcal{J}=\ell-\beta k$ of the conserved angular momentum, while the gauge holonomy of the pulled-back field, $\oint_{\mathcal C_n}qA=\pi q\mathcal B_0 r^2 n$ [Eq.~\eqref{eq:5.13}], supplies the confining scale rather than an independent Aharonov--Bohm parameter, since no additional flux threads the defect core in the present construction. Their combined effect is therefore not an interference phase but the orientation-dependent spectral asymmetry of Eq.~\eqref{eq:disc_spectral_asymmetry}, $\varepsilon_n(-\chi)-\varepsilon_n(\chi)=2\sigma\chi$: a spectroscopic, rather than interferometric, fingerprint of the same topological input identified in \cite{furtado1999landau,furtado2000berry,furtado2001scattering,delimaribeiro2003hmw}. In this sense the bound-state problem analyzed here complements the existing Aharonov--Bohm and Berry-phase treatments of the screw dislocation: it shows that, once the magnetic field confines the motion, the geometric holonomy responsible for those phase effects reorganizes the discrete spectrum itself, with the sign of $q\mathcal B_0 \mathcal{J}$ playing the same role that the sign of the enclosed flux plays in the scattering geometric phase.

The classical Hamiltonian therefore reduces to
\begin{equation}
\mathcal H_{\rm red}
=
\frac{p_r^2}{2\mu}
+\frac{k^2}{2\mu}
+\frac{(\mathcal{J}-ar^2)^2}{2\mu r^2},
\qquad
a=\frac{q\mathcal B_0}{2}.
\label{eq:disc_classical}
\end{equation}

Equation~(\ref{eq:disc_classical}) already exposes the basic physical
competition. At small radius,
\begin{equation}
V_{\rm eff}(r)\sim\frac{\mathcal{J}^2}{2\mu r^2},
\qquad r\rightarrow0^+,
\label{eq:disc_smallr}
\end{equation}
so that the screw-modified conserved momentum generates an inverse-square
barrier. At large radius,
\begin{equation}
V_{\rm eff}(r)\sim
\frac{q^2\mathcal B_0^2}{8\mu}r^2,
\qquad r\rightarrow\infty,
\label{eq:disc_larger}
\end{equation}
and the magnetic field supplies harmonic confinement. Thus the classical
motion is controlled by two distinct mechanisms: the global screw
identification determines the short-distance barrier, whereas the external
magnetic field fixes the large-distance confining scale. This separation is
also visible in the minimum of the dimensionless potential introduced
below.

Fixing the two cyclic momenta leaves a single radial degree of freedom.
Quantization of this reduced phase space gives a one-dimensional
Schr\"odinger problem on the half-line,
\begin{equation}
\left[
-\frac{1}{2\mu}\frac{d^2}{dr^2}
+\frac{k^2}{2\mu}
+\frac{\mathcal{J}^2}{2\mu r^2}
-\frac{q\mathcal B_0\mathcal{J}}{2\mu}
+\frac{q^2\mathcal B_0^2}{8\mu}r^2
\right]u
=
\mathcal E u .
\label{eq:disc_quantum}
\end{equation}
For \(q\mathcal B_0\neq0\), the inverse-square oscillator is exactly
solvable. Selecting the regular realization at the origin and imposing
normalizability at infinity reduces the confluent-hypergeometric solution to
an associated-Laguerre polynomial through the termination condition
\(\alpha=-n\). The resulting spectrum is
\begin{equation}
\mathcal E_n
=
\frac{k^2}{2\mu}
-\frac{q\mathcal B_0\mathcal{J}}{2\mu}
+
\frac{|q\mathcal B_0|}{2\mu}
\left[
2n+1+
\frac12\sqrt{1+4\mathcal{J}^2}
\right].
\label{eq:disc_exact_spectrum}
\end{equation}
Thus the exact quantum solution preserves the same separation already
identified classically: the screw geometry enters through \(\mathcal{J}\), whereas
the magnetic field supplies the confining scale and the oscillator ladder. This mechanism is the same qualitative effect identified for an electron in a uniform magnetic field in the presence of an isolated screw dislocation \cite{furtado1999landau} and in the presence of several coexisting defects \cite{marques2001landau}, as well as for a continuous distribution of screw dislocations \cite{silvanetto2008elastic}, in all of which the infinite Landau degeneracy is lifted by the topological boundary condition; here this lifting is obtained in closed form, together with the explicit localization scale \(\ell_B\) and the spectral-asymmetry relation of Eq.~\eqref{eq:7.31}.

The universal structure becomes particularly transparent after introducing
\begin{equation}
\chi=\mathcal{J},
\qquad
\sigma=\operatorname{sgn}(q\mathcal B_0),
\qquad
\ell_B=\sqrt{\frac{2}{|q\mathcal B_0|}},
\qquad
\rho=\frac{r}{\ell_B}.
\label{eq:disc_variables}
\end{equation}
The radial equation then takes the parameter-reduced form
\begin{equation}
\left[
-\frac{d^2}{d\rho^2}
+\frac{\chi^2}{\rho^2}
-2\sigma\chi+\rho^2
\right]u
=
2\varepsilon u ,
\label{eq:disc_universal}
\end{equation}
with
\begin{equation}
\varepsilon_n
=
-\sigma\chi
+2n+1
+\frac12\sqrt{1+4\chi^2}.
\label{eq:disc_universal_spectrum}
\end{equation}
Hence all local radial problems having the same \(\chi\) are dimensionless
equivalents, independently of the separate values of \(\ell\), \(k\), and
\(\beta\). The physical energy, however, still retains the axial kinetic
contribution \(k^2/(2\mu)\). This distinction between dimensionless radial
equivalence and full energetic equivalence is important when interpreting
the radial probability map in Fig.~\ref{fig:QR}(c).

The dimensionless effective potential,
\begin{equation}
\mathcal U_\chi(\rho)
=
\frac{\chi^2}{2\rho^2}
-\sigma\chi
+\frac{\rho^2}{2},
\label{eq:disc_U}
\end{equation}
has, for \(\chi\neq0\),
\begin{equation}
\rho_{\min}^2=|\chi|,
\qquad
\mathcal U_{\chi,\min}=|\chi|-\sigma\chi .
\label{eq:disc_Umin}
\end{equation}
The position of the minimum depends only on \(|\chi|\), whereas its energy
depends on the relative orientation of the magnetic field and the
screw-modified momentum. Panel~(a) of Fig.~\ref{fig:QR} displays the
resulting family of effective potentials, while panel~(b) shows the
corresponding spectral flow.

A particularly simple consequence is the different behavior under
\(\chi\rightarrow-\chi\). Since
\begin{equation}
s=\frac12\left(1+\sqrt{1+4\chi^2}\right)
\end{equation}
is even in \(\chi\), the exact Laguerre densities obey
\begin{equation}
|u_{n,\chi}(\rho)|^2
=
|u_{n,-\chi}(\rho)|^2,
\end{equation}
whereas the energies satisfy
\begin{equation}
\varepsilon_n(-\chi)-\varepsilon_n(\chi)
=
2\sigma\chi .
\label{eq:disc_spectral_asymmetry}
\end{equation}
The radial probability density is therefore insensitive to the sign of the
screw-modified momentum, whereas the spectrum retains this sign through the
magnetic coupling. This separation between the even radial dependence and
the sign-sensitive spectral term is a direct signature of the reduced
screw--magnetic coupling: the geometric contribution enters the centrifugal
sector through \(|\mathcal{J}|\), while its magnetic coupling produces the
spectral asymmetry. Panel~(d) shows the corresponding radial eigenfunctions
for the representative sector \(\chi=2\).

The eigenfunctions also determine the localization scale without
approximation,
\begin{equation}
\left\langle r^2\right\rangle_n
=
\frac{2}{|q\mathcal B_0|}
\left(
2n+s+\frac12
\right),
\label{eq:disc_mean_radius}
\end{equation}
and hence
\begin{equation}
\sqrt{\left\langle r^2\right\rangle_n}
\propto |q\mathcal B_0|^{-1/2}.
\label{eq:disc_localization}
\end{equation}
The magnetic field thus controls the absolute radial size through
\(\ell_B\), while \(|\chi|\) determines the dimensionless radial structure.
Increasing the field compresses the state in physical units without
altering its scaled profile. The dependence of the rms localization length
on \(\chi\) and the radial quantum number is shown explicitly in
Fig.~\ref{fig:QR}(e), while the transverse ground- and first-excited-state
densities are displayed in Fig.~\ref{fig:QR}(g) and (h), respectively.
The spectrum simultaneously forms an equally spaced ladder,
\begin{equation}
\varepsilon_{n+1}-\varepsilon_n=2,
\qquad
\mathcal E_{n+1}-\mathcal E_n
=
\frac{|q\mathcal B_0|}{\mu},
\label{eq:disc_spacing}
\end{equation}
showing that the screw coupling shifts the ladder but does not alter its
spacing.

The physical energies require an additional distinction because the
longitudinal kinetic contribution is not contained in the dimensionless
radial spectrum. In particular, two configurations can share the same
radial invariant \(\chi=\ell-\beta k\) while having different values of
\(k\), and therefore different physical energies through
\(k^2/(2\mu)\). The comparison shown in Fig.~\ref{fig:QR}(f) is constructed
with a common radial value \(\chi_{\rm common}=\chi_A=2.5\) in the numerical
evaluation, so that the separation between the two plotted energy ladders
comes from their distinct longitudinal kinetic terms. This panel therefore
concerns the relation between the dimensionless radial spectrum and the
full physical energy rather than a spectral flow in \(k\).

The zero-field sector must be treated separately. When
\(q\mathcal B_0=0\), the confining \(r^2\) term vanishes and
\begin{equation}
\mathcal H_0
=
-\frac{1}{2\mu}\frac{d^2}{dr^2}
+\frac{k^2}{2\mu}
+\frac{\mathcal{J}^2}{2\mu r^2}.
\label{eq:disc_zero}
\end{equation}
The system then becomes scale invariant on the half-line and no discrete
oscillator ladder is generated by the present mechanism. Thus the limit
\(q\mathcal B_0\rightarrow0\) is not merely a smooth removal of a small
spectral correction: it removes the confining scale itself and changes the
spectral character of the problem. The discrete spectrum
Eq.~(\ref{eq:disc_exact_spectrum}) should consequently be understood
strictly within the nonzero-field sector. An inverse-square-type interaction combined with a screw dislocation was previously examined in \cite{bakke2021inversesquare}; there the attractive inverse-square term originates from an induced electric dipole coupled to a radial electric field, whereas here it arises purely from the geometric screw-modified momentum \(\mathcal{J}\), with the confining scale supplied entirely by the external magnetic field rather than by the interaction itself.

The complete picture emerging from Fig.~\ref{fig:QR} is a separation of
roles among the geometric defect, the magnetic field, and the longitudinal
motion. The screw identification produces the conserved-momentum
combination \(\mathcal{J}=\ell-\beta k\), which determines the local radial
problem through \(\chi=\mathcal{J}\). The magnetic field supplies the
quadratic confining scale and introduces the orientation-dependent term
\(-\sigma\chi\), while the longitudinal motion contributes the independent
kinetic energy \(k^2/(2\mu)\) to the physical spectrum. Consequently, the
dimensionless radial densities depend on \(|\chi|\), the spectral branches
retain the sign of \(\chi\) through the magnetic coupling, and the physical
energy contains additional information through \(k\). These distinct
dependencies provide separate spectral and spatial signatures of the
screw-dislocation geometry and the applied magnetic field.

More broadly, the analysis demonstrates that nontrivial global geometry can
leave a direct quantum imprint even in the absence of local curvature.
Here the imprint is not an additional geometric potential but a
topologically induced coupling of conserved momenta. The exact solvability
then makes it possible to disentangle this global contribution from
magnetic confinement and to identify separately its effects on localization,
level spacing, and spectral asymmetry. This provides a compact framework
for probing global geometric defects through measurable radial and
spectroscopic responses.

%\section*{Acknowledgments}
%H.H. is grateful to Excellence project FoS UHK 2203/2025-2026 for the financial support.

\section*{ Credit authorship contribution statement}
\textbf{Claudio Furtado}: Conceptualization, Methodology, Formal Analysis, Writing - Original Draft, Investigation, Visualization, Writing - Review and Editing.\\
\textbf{Omar Mustafa}: Conceptualization, Methodology, Formal Analysis,  Writing - Original Draft, Investigation, Visualization, Writing - Review and Editing.\\
\textbf{Hassan Hassanabadi}: Conceptualization, Methodology, Formal Analysis, Writing - Original Draft, Investigation, Visualization, Writing - Review and Editing.\\
\textbf{Abdullah Guvendi}: Conceptualization, Methodology, Formal Analysis, Writing - Original Draft, Investigation, Visualization, Writing - Review and Editing.\\

\section*{Data availability}

This manuscript has no associated data.

\section*{Conflicts of interest statement}

No conflict of interest declared by the authors.

\section*{Funding}
 The work by C. Furtado is supported by the CNPq (project PQ Grant 1A No. 311781/2021-7).

\end{document}